\documentclass[aps,prb,twocolumn,superscriptaddress,footinbib]{revtex4-2}

\usepackage{graphicx}  
\usepackage{subfigure}
\usepackage{dcolumn} 
\usepackage{bm}
\usepackage{amssymb}   
\usepackage{comment}
\usepackage[colorlinks,linkcolor=blue,anchorcolor=blue,citecolor=blue,urlcolor=blue]{hyperref}
\usepackage{xcolor}
\usepackage{multirow}
\usepackage{amsmath}
\usepackage{mathrsfs}
\usepackage{mathcomp}
\usepackage{textcomp}
\usepackage{dsfont}
\usepackage{esint}
\usepackage{braket}
\usepackage{textgreek}
\usepackage{lipsum}
\usepackage{marvosym} 
\usepackage{centernot}
\usepackage{cancel}

\begin{document}
\preprint{APS/123-QED}

\title{Dissipative Berry phase effect in quantum tunneling}

\author{Xiao-Xiao Zhang}
\affiliation{Department of Physics and Astronomy \& Stewart Blusson Quantum Matter Institute, University of British Columbia, Vancouver, BC, V6T 1Z4 Canada}
\affiliation{RIKEN Center for Emergent Matter Science (CEMS), 2-1 Hirosawa, Wako, Saitama 351-0198, Japan}
\author{Naoto Nagaosa}
\affiliation{RIKEN Center for Emergent Matter Science (CEMS), 2-1 Hirosawa, Wako, Saitama 351-0198, Japan}
\affiliation{Department of Applied Physics, The University of Tokyo, 7-3-1 Hongo, Bunkyo-ku, Tokyo 113-8656, Japan}


\newcommand\dd{\mathrm{d}}
\newcommand\ii{\mathrm{i}}
\newcommand\ee{\mathrm{e}}
\newcommand\zz{\mathtt{z}}
\makeatletter
\let\newtitle\@title
\let\newauthor\@author
\def\ExtendSymbol#1#2#3#4#5{\ext@arrow 0099{\arrowfill@#1#2#3}{#4}{#5}}
\newcommand\LongEqual[2][]{\ExtendSymbol{=}{=}{=}{#1}{#2}}
\newcommand\LongArrow[2][]{\ExtendSymbol{-}{-}{\rightarrow}{#1}{#2}}
\newcommand{\cev}[1]{\reflectbox{\ensuremath{\vec{\reflectbox{\ensuremath{#1}}}}}}
\newcommand{\red}[1]{\textcolor{red}{#1}} 
\newcommand{\blue}[1]{\textcolor{blue}{#1}} 
\newcommand{\green}[1]{\textcolor{orange}{#1}} 
\newcommand{\mycomment}[1]{} 
\makeatother

\begin{abstract}
Berry phase effect plays a central role in many mesoscale condensed matter and quantum chemical systems that are naturally under the environmental influence of dissipation. We propose and microscopically derive a prototypical quantum coherent tunneling model around a monopole or conical potential intersection in order to address the intriguing but overlooked interplay between dissipation and topologically nontrivial Berry phase effect. 
We adopt the instanton approach with both symmetry analysis and accurate numerical solutions that consistently incorporate nonperturbative dissipation and Berry phase. It reveals a novel dissipative quantum interference phenomenon with Berry phase effect. The phase diagram of this tunneling exhibits Kramers degeneracy, nonmonotonic dependence on dissipation and a generic dissipation-driven phase transition of quantum interference, before which an unconventional dissipation-enhanced regime of quantum tunneling persists.
\end{abstract}


\maketitle


\section{Introduction}
The effect of Berry phase as the intrinsic leading quantum correction in $\hbar$ resides at the heart of plenty of 
phenomena\cite{Berry1984,Bohm2003,Xiao2010}. It includes the intrinsic or quantized Hall transport in electronic, spin, thermal and other channels\cite{Xiao2010,Nagaosa2010,Sinova2015}, as well as the various electrical and optical manipulation of low-dimensional quantum materials\cite{\mycomment{Xiao2007,}Xiao2012,*Cao2012,*Xu2014,DiracFermion2}, and the expanding family of topological materials\cite{Topo1,*Topo2,ReviewYan2017,*WeylDiracReview}. Those phenomena are of both theoretical and practical significance partially because they connect the subatomic realm of quantum mechanical phase coherence and interference to potential applications in functional quantum materials at microscales and mesoscales\cite{Basov2017,*Tokura2017,*Keimer2017}. An immediate but underaddressed issue is that such systems inevitably are open and subject to \mycomment{decoherence and }dissipation due to the influence from the omnipresent environmental influence\cite{Zurek2003,Dattagupta2004,Weiss2012,Caldeira2014}.

A natural question arises: how would quantum dissipation affect Berry phase and will their interplay lead to any unique phenomena? To properly address the question, quantum geometric phases and dissipation need to be considered \textit{together}, which remains largely unexplored so far. Along the system-plus-reservoir way to quantum dissipation\cite{Caldeira1981,*Caldeira1983,Leggett1987,Weiss2012,Caldeira2014}, a powerful approach 
is, 
to the leading order of $\hbar$, the instanton\cite{Coleman1977,*Callan1977,*Coleman1985}. Without dissipation, instantons can crucially carry Berry phase and accommodate interference
, e.g., in Josephson junctions and micromagnets\mycomment{microscale magnetic grains or domains}\cite{\mycomment{Garg1993,}Loss1992,*Delft1992,*Braun1996,*\mycomment{Garg1999,}Leuenberger2001}, quantum spin chains\cite{Affleck1986} and quantum phase transitions\cite{Read1989,*\mycomment{Read1990,}Senthil2004}. 
However, dissipative instantons with Berry phase have been overlooked. Firstly, dissipation has been mostly considered for the interaction between instantons separated in time in an instanton gas, i.e., the \textit{inter}-instanton effect, where the Berry phase scarcely can enter\cite{Bray1982,Schmid1983,Leggett1987}. Secondly, the \textit{intra}-instanton effect of dissipation on each instanton itself, partially due to analytical difficulty, is often ignored\cite{Schmid1983,Callan1992} or included only in vanishing or overdamped limits\cite{Garg1989,*Garg1993a,*Garg1994} and for topologically trivial one-dimensional potentials\cite{Chang1984,Grabert1987}. Conversely, this part of dissipation on each instanton, as it strongly affects the tunneling trajectory, exactly suggests that it must have nonnegligible effects on the Berry phase accumulated along each instanton path.

\begin{figure}[hbt]
\includegraphics[width=8.6cm]{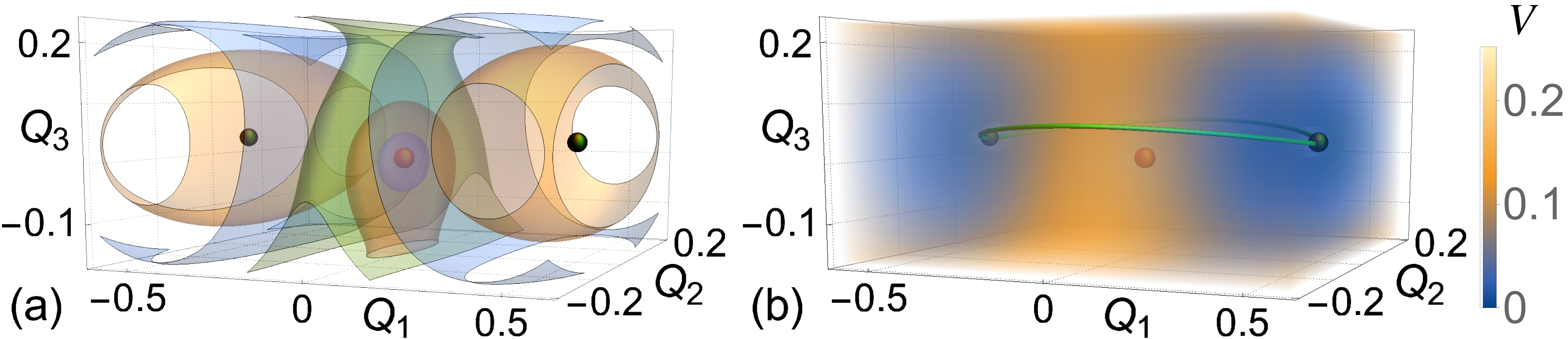}
\caption{Illustration of the potential $V(\vec{Q})$ for (coherent tunneling) CT with two black degenerate minima and a red monopole. (a) Potential contour surfaces equally spaced in energy. (b) Color-scaled potential plot with a typical pair of CT instanton paths connecting two minima.}
\label{Fig:V}
\end{figure}
Here, to address this issue in a systematic way accounting for the Berry phase and dissipation \textit{at once}, we will focus on a prototypical tunneling problem in a three-dimensional 
potential $V(\vec{Q})$ with a singular monopole structure and potential minima. 
Among two primary types of quantum tunneling processes, the quantum decay (QD) of a metastable state and the coherent tunneling (CT) between two degenerate minima, only CT, illustrated in Fig.~\ref{Fig:V}, will be relevant as to be explained. 
The monopole as the simplest topologically nontrivial source of Berry gauge field \mycomment{characterized by the second Homotopy group }is essential for a wide range of phenomena, since it generates the topological winding of emergent magnetic monopole in real space\cite{Goddard1978,*spinice0\mycomment{,*EEMF2},*XXZ:resistivity} and of degeneracy or band-crossing in momentum and parameter space of most topological insulators, superconductors, and (semi)metals\cite{Topo1,Topo2,ReviewYan2017,WeylDiracReview,DiracFermion1}. This monopole enters our model as a two-band conical potential intersection that can alternatively has its origin in Jahn-Teller systems\cite{\mycomment{Herzberg1963,*Obrien1964,}Khomskii1982} and many other scenarios in molecular physics and quantum chemistry\cite{Yarkony1996,*Domcke2004} as a phase singularity of 
molecular wavefunctions. As tunneling events of electrons and atoms between potential minima around an intersection form a typical reactive configuration in inorganic and even organic photochemical reactions, our study provides the missing information of dissipative Berry phase effects, which can alter the detectable reaction rate and scattering cross section in rate constants, kinetic isotope effects, etc\cite{Haenggi1990,Kryvohuz2011,*Richardson2011,Child2014,Meisner2016}.
\section{Model system}\label{Sec:model}
The motivation outlined above is embodied in the imaginary-time action
\begin{equation}\label{eq:S_final_main}
\begin{split}
    \mathcal{S} =   \mathcal{S}_\mathrm{\Phi} + \mathcal{S}_\mathrm{Q}   & = \mathcal{S}_\mathrm{\Phi}+ \int_{-T}^{T} \dd\tau [\sum_i \frac{1}{2} M_i\dot{Q}_i^2(\tau) + V(\vec Q) \\
    & + \sum_i \int_{-T}^{T}{ \dd\tau' \frac{D_i}{4} \frac{(Q_i(\tau)-Q_i(\tau'))^2}{(\tau-\tau')^2}  } ] 
\end{split}
\end{equation}
that describes a reactive quasiparticle or collective mode $\vec{Q}=(Q_1,Q_2,Q_3)$ with dissipative temporally nonlocal self-interaction, where mass $M_i$ and dissipation strength $D_i$ can in general be anisotropic. The quasiparticle feels a reversed potential $-V(\vec{Q})$ with
\begin{equation}\label{eq:potential_main}
    {V}(\vec{{Q}}) \mycomment{= \sum_{i=1,2,3}{\mu_i q_i^2(\tau)} -|\vec Q|} =  \sum_{i=1,2,3}{ ({Q}_i - {w}_i)^2 / {\alpha}_i - |\vec{{Q}}| }
\end{equation}
consisting of harmonic confinement and the lower energy surface of a conical intersection at $\vec{Q}=\vec{0}$. 
Anisotropic $\alpha_i$ and potential offset $w_i$, seen later, help form potential minima around the monopole (used interchangeably with conical intersection). 
The monopole Berry phase $\Phi$ attached to the quasiparticle is 
\begin{equation}\label{eq:spinBerry_main}
\mathcal{S}_\mathrm{\Phi} = \ii \Phi = \ii S \int_{-T}^T{\dd\tau (1-\cos\theta)\dot\phi}
\end{equation}
where $S$ as a factor dependent on specific systems is spin-\textrm{\textonehalf} per the construction below. The polar and azimuthal angles $\theta,\phi$ in $\vec{Q}$-space satisfy $\cos\theta=Q_3/|\vec{Q}|$ and the one-form $\dot\phi=(Q_1\dot Q_2-Q_2\dot Q_1)/(Q_1^2+Q_2^2)$, which is made only singular along northbound or southbound Dirac strings of the monopole. In contrast to one-dimensional instanton trajectories often with no phase accumulation and two-dimensional quadratic Aharonov-Bohm phase terms under magnetic field\cite{Callan1992}, the highly nonlinear dependence on $\vec Q$ in Eq.~\eqref{eq:spinBerry_main} reflects the nontrivial topological covering of the $S^2$-sphere due to the monopole. 

Although this phenomenological model stems from foregoing motivations and could also derive from 
a suitable type of semiclassical wave-packet dynamics\cite{Xiao2010}, 
it is still informative to place a viable microscopic basis.
Suggested by spin Berry phase's similar form\mycomment{in coherent-state path integral} as Eq.~\eqref{eq:spinBerry_main}, quantum spin-$S$ bears the same dynamics as a massless charge $S$ under monopole gauge field\cite{CMFT,EEMF2}. For instance, simple (pseudo)spin system $H_\mathrm{s} = \vec{w}\cdot\vec{\sigma}$ holds a Berry monopole at $\vec{w}=\vec{0}$, which can represent, e.g., a spin under magnetic field and Weyl semimetal bands when $\vec{w}$ respectively stands for magnetic field and momentum.
Exploiting this connection, we can motivate and derive Eq.~\eqref{eq:S_final_main} from a modified spin-boson model
\begin{equation}\label{eq:H_all_main}
    H = H_\mathrm{s} + H_\mathrm{b} + H_\mathrm{c}\mycomment{ + H_\mathrm{ct}}
\end{equation}
describing a spin 
couples via $H_\mathrm{c} = \sum_i{c_i \sigma_i q_i}$ to a group of harmonic oscillators 
$H_\mathrm{b} = \sum_{i\nu}{\frac{m_\nu}{2}(\dot x_{i\nu}^2 + \omega_\nu^2x_{i\nu}^2)}$
labeled by $\nu$ and direction $i$,
where the collective coordinate $q_i=\sum_\nu{g_{i\nu}x_{i\nu}}$ more of macroscopic nature is specified by \mycomment{
$\{g_{i\nu}\}$ determined by }particular systems.\mycomment{One can integrate out the boson fields except the collective mode $\vec{q}$, resulting in an effective action in Matsubara frequency $\omega$
\cite{SM}
\begin{equation}\label{eq:S_eff_main}
    \mathcal{S}_\mathrm{eff} = \mathcal{S}_\mathrm{s} +  2T\sum_{i,\omega}{ ( u_{i\omega}^{-1} q_{i\omega}^* q_{i\omega}  + c_i\sigma_{i\omega}^*q_{i\omega} ) } 
\end{equation}
where the spin action $\mathcal{S}_\mathrm{s} = \mathcal{S}_\mathrm{\Phi} + \int_{-T}^{T}{\dd \tau H_\mathrm{s}}$ with spin Berry phase $\mathcal{S}_\mathrm{\Phi}$ in terms of Bloch-sphere angles formally identical to Eq.~\eqref{eq:spinBerry_main}. $u_{i\omega} = \frac{4}{\pi c_i^2} \int{ \dd\omega' J_i(\omega') \frac{\omega'}{\omega^2+\omega'^2} }$ with the coupling spectral density $J_i(\omega) \mycomment{= \frac{\pi}{2} \sum_\nu{ \frac{c_i^2g_{i\nu}^2}{m_\nu\omega_\nu} \delta(\omega-\omega_\nu) }}$. $J_i$ in reality differs from the memoryless ohmic friction $J_0(\omega) = \eta\omega$ and decays fast enough as, for instance, an analytically tractable Lorentzian-like regularization does $J_i(\omega) = \eta_i\omega\mycomment{^s}/(1+\omega^2/\omega_\mathrm{D}^2)^2$, 
\mycomment{which is ohmic for $s=1$ }and acquires a memory-friction time scale $1/\omega_\mathrm{D}$ \blue{in the damping function\mycomment{ in generalized Langevin equations}}. 
}
Once we identify the transform $Q_i = w_i + c_iq_i$ that conveniently places the monopole at the origin $\vec{Q}=\vec{0}$, Eq.~\eqref{eq:S_final_main} with $M_i=\frac{1}{\alpha_i},D_i=\frac{d}{\alpha_i}$ can be generated as the leading contribution from integrating out the environment except the collective mode $\vec{Q}$, as detailed in Append.~\ref{App:Derivation}. 
In accordance with Jahn-Teller potential energy surfaces, we take the adiabatic approximation as spin $\vec\sigma$ inclines to follow the more macroscopic variable $\vec{Q}$\cite{Leggett1997} and the $\vec{Q}$-motion hence inherits the spin Berry phase as Eq.~\eqref{eq:spinBerry_main}.
As all quantities are nondimensionalized, we have dimensionless potential parameters $\vec\alpha,\vec w$ and dissipation $d$ henceforth.

\mycomment{One also identifies a Hamiltonian $H_0 = \mycomment{H_\mathrm{s} + H_\mathrm{c} =  }\vec{Q}\cdot\vec\sigma$ from the rest in Eq.~\eqref{eq:S_eff_main}. In accordance with Jahn-Teller potential energy surfaces, we take the adiabatic approximation as $\vec\sigma$ inclines to follow the more macroscopic variable $\vec{Q}$\cite{Leggett1997}. $H_0$ is replaced by energy $-|\vec{Q}|$ in Eq.~\eqref{eq:potential_main} and $\vec{Q}$ is manifestly the variable most relevant to the dynamics. Now the motion in $\vec{Q}$ inherits the spin Berry phase\mycomment{ as Eq.~\eqref{eq:spinBerry_main}}, which we cast as
$
    \cos\theta=Q_3/|\vec{Q}|\,,\dot\phi=(Q_1\dot Q_2-Q_2\dot Q_1)/(Q_1^2+Q_2^2)$.
The \mycomment{closed but nonexact }one-form $\dot\phi$ facilitates 
evaluation since it is only singular along \mycomment{the northbound or southbound }polar Dirac strings.
}
The emergence of monopole phase or Weyl-like potential 
originates from that spin-$S$ bears the dynamics of a massless charge $S$ under a \mycomment{$\mathrm{U}(1)$ }monopole gauge field $\vec{A}$. In fact, $\mathcal{S}_\mathrm{\Phi}$ can be written as an orbital $\vec{j}\cdot\vec{A}$-type coupling $\ii S \int_\gamma \dd\hat{n}\cdot\vec{A}$ where 
$\hat{n}=\vec{S}/|\vec{S}|$ has trajectory $\gamma$\mycomment{ and for instance, $\vec{A}=\frac{1-\cos\theta}{\sin\theta}\hat{e}_\phi$ has a southbound Dirac string}. 
The mechanism lies in $\mathrm{SU}(2) \cong S^2 \times \mathrm{U}(1)$ of spin $\mathrm{SU}(2)$ and monopole manifold $S^2$. 
Therefore, 
we are finally led to
Eq.~\eqref{eq:S_final_main}.

\section{Instanton equation and potential landscape}\label{Sec:EOM}
For Eq.~\eqref{eq:S_final_main}, we will use the instanton technique to include the leading nonperturbative effect in the exponential contribution to the transition amplitude
$
    \langle \vec{Q}(T)\lvert \ee^{-2\mathcal{H}T} \rvert\vec{Q}(-T)\rangle=\int\mycomment{\limits_{\vec{Q}(-T)}^{\vec{Q}(T)}} \mathcal{D} \vec{Q}(\tau) \ee^{-\mathcal{S}[\vec{Q}(\tau)]}
$
close to zero temperature with 
large enough $T$. 
The fluctuation upon instantons accounting for the less dominant non-exponential prefactor is not considered here\mycomment{\cite{Garg1992}}. As emphasized earlier, the nontrivial dissipative Berry phase only enters through the intra-instanton dissipation, where instanton trajectories and accumulated Berry phases consistently depend on dissipation. It has seldom been treated appropriately and the common approximation using the dissipationless instanton instead can simply lose dissipative Berry phase.
As per the adiabatic approximation and the original realness of coordinate $\vec{Q}$, the correct way is to truly solve for 
$\mathcal{S}_\mathrm{Q}$ in the presence of finite dissipation while the imaginary $\mathcal{S}_\mathrm{\Phi}$ attaches a dissipation-dependent complex phase to the quantum amplitude, i.e., we need the genuine dissipative instanton with Berry phase.

\begin{figure}[hbt]
\includegraphics[width=8.6cm]{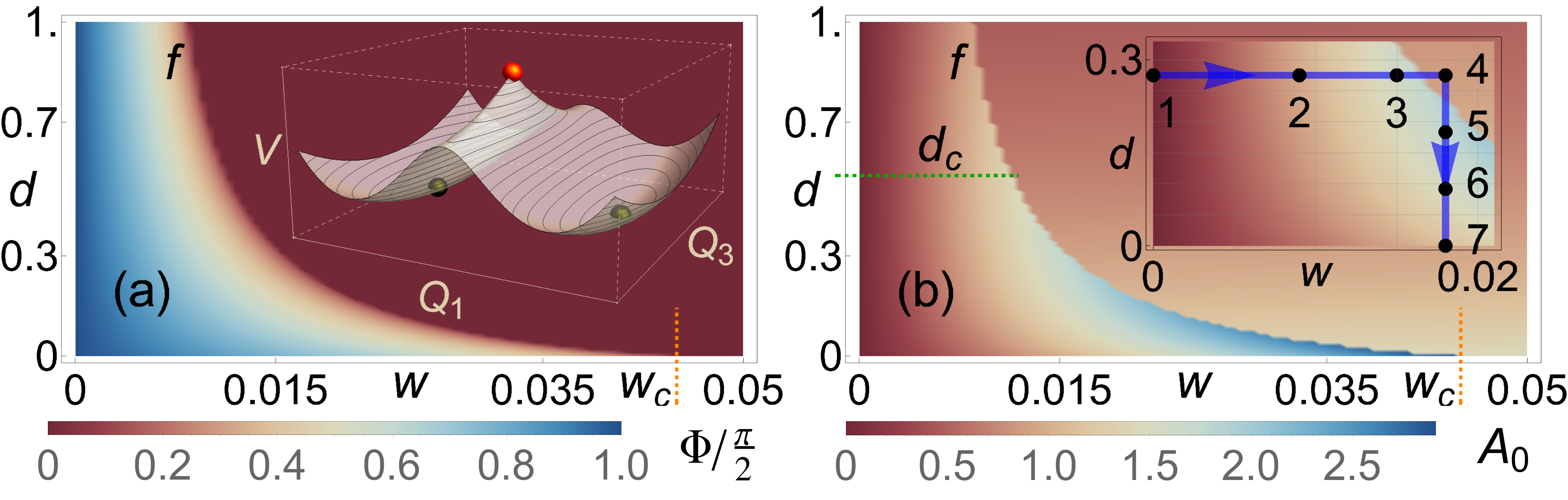}
\caption{Phase diagrams of the dissipative CT's (a) instanton Berry phase $\Phi$ and (b) transition amplitude $A_0$ against potential bias $w$ and dissipation $d$. The phase boundary lies at the critical curve $f$ where interference disappears. Dashed lines: orange, maximal $w_c$ on $f$; green, upper $d_c$ of the dissipation-enhanced anomalous $A_0$. Inset(a): potential of Fig.~\ref{Fig:V} plotted in the $Q_2=0$-plane. Inset(b): seven points along the arrowed path are used in Fig.~\ref{Fig:Paths}. Instantons are solved on a 200-point $\tau$-grid with parameters $\mycomment{S=\frac{1}{2},}2T=10,\vec{\alpha}=(1,0.6,0.4)$, scanning a $100\times 100$-mesh of the $wd$-plane.}
\label{Fig:PhaseDiagram}
\end{figure}

The full instanton equation reads
\begin{equation}\label{eq:EOM_main}
{M}_i\frac{\dd^2{Q}_i}{{\dd x}^2} - \frac{\partial{V}(\vec{{Q}})}{\partial {Q}_i} 
- {D}_i \int_{- T}^{ T}{ \dd \tau' \frac{{Q}_{i}(\tau)-{Q}_{i}(\tau')}{(\tau-\tau')^2} }  = 0.
\end{equation}
Its boundary conditions (BCs) for QD and CT are respectively $\vec{Q}(\pm T)=\vec{Q}_0$ and $\vec{Q}(\pm T)=\vec{Q}_\pm$ where $V(\vec{Q}_0)$ is a metastable minimum and $V(\vec{Q}_\pm)$ signify two degenerate minima. The anisotropy in $\vec{\alpha},\vec{w}$ and the mirror symmetry $\mathcal{M}_i$ of $V(\vec{Q})$ under $Q_i\rightarrow -Q_i$ when $w_i=0$
help create corresponding potentials. To determine the extremum manifold, 
we inspect the \mycomment{principal minors }positive-definiteness of the Hessian $\frac{\partial^2 V}{\partial Q_i \partial Q_j}$ in Append.~\ref{App:landscape}. 
Explained shortly, QD instantons have no Berry phase interference, we therefore mostly work on representative CT cases assuming\mycomment{ $\vec w$ is aligned with one axis,} $\alpha_1>\alpha_2>\alpha_3$: 
$\vec{w}=w\hat{3},\vec Q_\pm = (\pm\alpha_1\sqrt{\frac{1}{4}-(\frac{w}{\alpha_1-\alpha_3})^2}, 0, \frac{\alpha_{1}w}{\alpha_{1}-\alpha_3})$ when $\alpha_1-\alpha_3>2w$, illustrated in Fig.~\ref{Fig:V} and Fig.~\ref{Fig:PhaseDiagram}(a) inset; and the generalization to $\vec{w}=w_2\hat{2}+w_3\hat{3}$.



We have a difficult boundary-value problem (BVP) of a system of nonlinear integro-differential equations with a Fredholm integral. Its lack of systematic treatment and nonunique solutions partially obstructs early studies. Certain related methods applied here
could be uncontrolled or require exceeding computational cost towards correct convergence\cite{Arikoglu2005,Trefethen2017,He2006}. 
We instead propose a viable and generalizable approach widely applicable to dissipative instanton problems, which is detailed in Append.~\ref{App:FDM}. Its essence is, in the spirit of finite difference method (FDM)\cite{Leveque2007,Trefethen2017}, using irregular FDM stencils\cite{Fornberg1988} generated from Gaussian quadrature rules\cite{AbramowitzStegun} to convert simultaneously the differential and integral parts to a system of nonlinear algebraic equations, which can be solved by root-finding algorithms with initial guesses, e.g., Newton's and Brent's hybrid methods\cite{NumericalRecipes,GSL}.
Finer solutions are readily facilitated using existing coarse and/or nearby-in-parameter solutions as initial warm-ups. This not only accelerates solutions but particularly helps reach symmetry-related instantons at will. 
All the properties discussed below are verified numerically in extensive parameter choices.

\begin{table}
\begin{tabular}{c|c|c}
& $\mathrm{loss}$ & $\mathrm{gain}$\\
\hline
$\mathcal{S}_\mathrm{Q}$ & $\vec{Q}(\tau),\vec{Q}_\mathrm{h}\rightarrow \vec{Q}_\mathrm{l}$ & $\mathcal{T}\vec{Q}(\tau),\vec{Q}_\mathrm{l}\rightarrow \vec{Q}_\mathrm{h}$\\
\hline
$\mathcal{S}_\mathrm{Q}'(\geq\mathcal{S}_\mathrm{Q})$ & $\mathcal{T}\vec{Q}'(\tau),\vec{Q}_\mathrm{l}\rightarrow \vec{Q}_\mathrm{h}$ & $\vec{Q}'(\tau),\vec{Q}_\mathrm{h}\rightarrow \vec{Q}_\mathrm{l}$\\
\end{tabular}
\caption{Action and energy behavior of two generic solution paths $\vec{Q}(\tau),\vec{Q}'(\tau)$ and their time-reversal. They connect $\vec{Q}_\mathrm{h,l}$ with potential $V_\mathrm{h}\geq V_\mathrm{l}$.}\label{table1}
\end{table}
\section{Symmetry-related dissipative instantons with Berry phase}\label{Sec:symmetry}
Although we can calculate the full dissipative instantons, it is crucial to understand analytically the symmetry properties under dissipation beforehand. For a generic BVP of Eq.~\eqref{eq:EOM_main} connecting two points or local minima $\vec{Q}_\mathrm{h,l}$ with $-V_\mathrm{h}\leq -V_\mathrm{l}$ in reversed potential, which is relevant to both QD and CT, one always finds two valid solutions: $\vec{Q}(\tau)$ of globally minimal action $\mathcal{S}_\mathrm{Q}$ exhibits net energy loss; another $\vec{Q}'(\tau)$ of locally minimal action $\mathcal{S}_\mathrm{Q}'\geq\mathcal{S}_\mathrm{Q}$ exhibits net energy gain, because dissipating initial kinetic energy to travel through $\vec{Q}_\mathrm{h}\rightarrow\vec{Q}_\mathrm{l}$ is faster than gradually absorbing energy to accelerate. Summarized in Table~\ref{table1}, they also have respective time-reversed ($\mathcal{T}$) counterparts of opposite energy variation that share the same action (see also Append.~\ref{App:irreversibility}).
This is because although the microscopic time-reversal symmetry $\mathscr{T}$ is broken by Zeeman-like $\vec{w}$ in Eq.~\eqref{eq:H_all_main}, contrary to some common misconception, the $\tau$-reversibility $\mathcal{T}$ in Eqs.~\eqref{eq:S_final_main}\eqref{eq:H_all_main} remains intact. 
Temporally retarded or advanced self-interaction can both be induced from exchanging fluctuations with environment: $\vec{Q}(\tau)$ and $\mathcal{T}\vec{Q}(\tau)=\vec{Q}(-\tau)$ of interchanged BCs are both valid paths satisfying Eq.~\eqref{eq:EOM_main} since the integro-differential operator is \textit{parity even}, different from viscous Newtonian or Langevin equations that solely possess lossy solutions.
The irreversibility in those lossy equations derives from both densely distributed many-body states \mycomment{at thermodynamic limit }due to the environment and the causal structure \mycomment{in subsystem's self-energy }from infinitesimal 
relaxation\cite{Datta2005,Kamenev2009}. 
Also, shown in Append.~\ref{App:QD}, one can use Table~\ref{table1} to infer that any dissipative QD instanton does not carry Berry phase because it is energetically determined to be $\mathcal{T}$-symmetric.
The relevance of $\mathcal{T}$-paths together with the nonuniqueness of solution (see Table~\ref{table2}) has long been disregarded, hiding the rich possibility of dissipative Berry phase effects.

\begin{table}
\begin{tabular}{c|c|c}
mirror symmetry & instanton & phase\\
\hline
\multirow{2}{9em}{$\mathcal{M}_1:\vec{w}=w_2\hat{2}+w_3\hat{3}$}
& $\vec{Q}$ & $\Phi$ \\
& $\vec{Q}'=\mathcal{M}_1\mathcal{T}\vec{Q}$ & $\Phi'=\Phi$\\
\hline
\multirow{4}{8em}{$\mathcal{M}_{1,2}:\vec{w}=w\hat{3}$}  
& $\vec{Q}$ & $\Phi$ \\
& $\vec{Q}'=\mathcal{M}_1\mathcal{T}\vec{Q}$ & $\Phi'=\Phi$ \\
& $\vec{Q}^r=\mathcal{M}_2 \vec{Q}$ & $\Phi^{\mathrm{r}}=-\Phi$\\
& $\vec{Q}^{\mathrm{r}\prime}=\mathcal{M}_2 \mathcal{M}_1\mathcal{T}\vec{Q}$ & $\Phi^{\mathrm{r}\prime}=-\Phi$\\
\end{tabular}
\caption{Mirror symmetry condition and valid dissipative instantons with Berry phases that are related to a generic CT instanton $\vec{Q}(\tau)$ with phase $\Phi$.}\label{table2}
\end{table}
Let's henceforth focus on CT corresponding to the $V_\mathrm{h}=V_\mathrm{l},\mathcal{S}_\mathrm{Q}'=\mathcal{S}_\mathrm{Q}$ case of Table~\ref{table1}, where $V(\vec{Q})$ requires at least one emergent mirror symmetry and $\vec{Q}'=\mathcal{M}_1\mathcal{T}\vec{Q}$. As detailed in Append.~\ref{App:mirror}, the key observation at finite dissipation is twofold. Firstly, dependent on the number of mirror symmetries, using Eqs.~\eqref{eq:S_final_main}\eqref{eq:spinBerry_main}\eqref{eq:EOM_main} and the parity of 
BCs $Q_i(-T)=\pm Q_i(T)$, one finds in Table~\ref{table2} at most three extra symmetry-related instantons of the same $\mathcal{S}_\mathrm{Q}[\vec{Q}(\tau)]$ as $\vec{Q}$: $\vec{Q}'$ aforementioned and a new $\vec{Q}^\mathrm{r}$ with its own $\vec{Q}^{\mathrm{r}\prime}$. 
Secondly, instantons $\vec{Q}^\mathrm{r},\vec{Q}^{\mathrm{r}\prime}$ carry a reversed Berry phase $\Phi^{\mathrm{r}(\prime)}=-\Phi$ and thus collectively interfere with $\vec{Q},\vec{Q}'$ as long as $\Phi\neq0$.
Note that one cannot arbitrarily translate a system to accommodate $Q_2(\pm T)=0$ and obtain $\vec{Q}^{\mathrm{r}(\prime)}$ instantons since the two parts in $V(\vec{Q})$ have different inversion centers. Since we have consistently incorporated dissipation to instanton trajectories and associated Berry phases, we are ready to combine multiple instantons towards the transition amplitude under dissipation
\begin{equation}\label{eq:CT_main}
    A_0 = \textstyle \sum_i \ee^{-\mathcal{S}_{\mathrm{Q}}[\vec{Q}^{(i)}]-\ii\,\Phi[\vec{Q}^{(i)}]} 
    = 4\ee^{-\mathcal{S}_\mathrm{Q}}  \cos{\Phi},
\end{equation}
where $i$ runs over all four instantons in the second row of Table~\ref{table2} because any of them, loss or gain, should contribute in the path integral. 
As long as coherence persists, the probability of finding the system at one minimum oscillates at a frequency of the energy splitting, $\Delta\propto A_0$ discussed later, between spontaneously formed even- and odd-parity states. 

\section{Kramers degeneracy, phase transition of quantum interference, and dissipation-enhanced CT}\label{Sec:phasediagram}
Based on Eq.~\eqref{eq:CT_main} from symmetry analysis, now we study the concrete dependence on dissipation $d$ and potential bias $\vec{w}=w\hat{3}$ of the more interesting second CT case in Table~\ref{table2} with Berry phase interference. We mainly exemplify with the $\vec{Q},\vec{Q}^\mathrm{r}$ instantons, representing as well the similar and $\mathcal{M}_1\mathcal{T}$-symmetry-related behavior of the $\vec{Q}^\prime,\vec{Q}^{\mathrm{r}\prime}$ pair.
Importantly, including Berry phase consistently under nonperturbative dissipation enables us to find the rich variation of $\Phi,A_0$ in Fig.~\ref{Fig:PhaseDiagram}, which has been ignored and would be totally absent if the intra-instanton dissipation effect on Berry phase were not at play.

\begin{figure}[t]
\includegraphics[width=8.6cm]{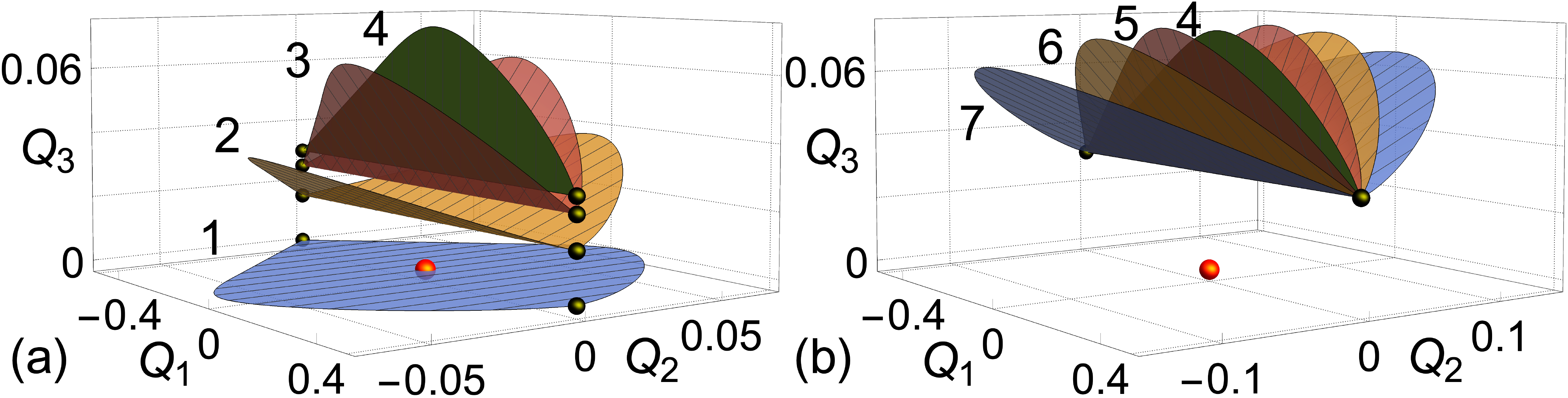}
\caption{Evolution of the representative $\mathcal{M}_2$-symmetry-related  $\vec{Q},\vec{Q}^\mathrm{r}$ instanton pair along the path in Fig.~\ref{Fig:PhaseDiagram}(b) inset, which (a) increases $w$ until case '4' outside the phase boundary (the merged single green instanton) and then (b) decreases $d$ down to zero. Another distinct instanton pair $\vec{Q}^\prime,\vec{Q}^{\mathrm{r}\prime}$ with the similar behavior are not shown. Black and red dots follow those of Fig.~\ref{Fig:V} displaying the typical potential.
}
\label{Fig:Paths}
\end{figure}

In Fig.~\ref{Fig:PhaseDiagram}(a), Berry phase $\Phi$ 
decreases with $w$, because at larger $w$ instantons travel higher above the monopole and thus acquires less flux as shown in Fig.~\ref{Fig:Paths}(a). Similarly corresponding to Fig.~\ref{Fig:Paths}(b), $\Phi$ follows the variation of $\max[|Q_2(\tau)|]$ with $d$ since $Q_2$ is the mirror-symmetric direction expanding solid angle against the monopole. 
Interestingly, along $w=0$, any dissipation accommodating finite $\max[|Q_2(\tau)|]$ attains a hemisphere covering of the monopole and hence the topological prohibition of tunneling $A_0=0$ at $\Phi=\pi/2$ in Fig.~\ref{Fig:PhaseDiagram}(b). Recalling the microscopic model Eq.~\eqref{eq:H_all_main} with spin, this reflects the original Kramers degeneracy even under dissipation as the microscopic time-reversal $\mathscr{T}$ is not broken when $w=0$ and $\vec{Q},\vec{Q}^\mathrm{r}$ instantons interfere destructively ['1' in Fig.~\ref{Fig:Paths}(a)].

More significantly, a novel phase transition 
of dissipative Berry phase interference occurs across the phase boundary $f$ in Fig.~\ref{Fig:PhaseDiagram}, at which $\Phi$ continuously vanishes and is singular in its derivative while $A_0$ or $\Delta$ discontinuously jumps downwards. 
We illustrate the physical picture of the evolution and the transition in Fig.~\ref{Fig:Paths}. 
For instance, from '1' to '4' in Fig.~\ref{Fig:Paths}(a), the $\mathcal{M}_2$-symmetric $\vec{Q},\vec{Q}^\mathrm{r}$ instanton pair at $w=0$ is inside the $Q_1Q_2$-plane, and, as $w$ increases, they\mycomment{, in general in 3D due to the nonconservation of angular momentum from $V$,} distort and bend towards each other until coalescence at $f$, i.e., merging into a single green instanton in the $Q_1Q_3$-plane.
Hence, the number of contributing instantons is halved in Eq.~\eqref{eq:CT_main} and no interference outside $f$. We note that the discontinuity along $f$ 
can become a sharp crossover, because enhanced quantum fluctuations of the typical scale\cite{Schulman2005} $|\hbar/\partial_{\vec{Q}}^2\mathcal{S}|^\frac{1}{2}$ will prevail around the instanton and smear two merging instantons separated within this scale in $\vec{Q}$-space, where the independent instanton viewpoint falls short. As $\hbar\ll\mathcal{S}$ for general mesoscopic systems, this crossover is narrow and the sharp transition remains well observable in tunneling probability, degeneracy splitting, reaction rate, etc.
This unconventional transition of Berry phase interference in CT, driven by the mechanism of instanton merging under dissipation and monopole potential, displays a new generic type of phenomenon of how quantum phase coherence and interference survives dissipation until a critical strength. It is fundamentally distinct from the dissipation-induced localization transition to be mentioned shortly and is never possible until consistent inclusion of intra-instanton dissipation with Berry phase. We also give more analytical analysis and discuss the large-$d$ behavior in Append.~\ref{App:PhaseDiagram}.

Dissipation is conventionally simply reckoned to suppress quantum effects 
as the environment constantly measures and renders the system classical. 
$\mathcal{S}_\mathrm{Q}$ in our calculation indeed increases with $d$ partly due to the positive nonlocal interaction in Eq.~\eqref{eq:S_final_main}, which underlies the common belief of dissipation-suppressed CTs. 
Yet another surprising consequence of the dissipative Berry phase is the nonmonotonic variation in Fig.~\ref{Fig:PhaseDiagram}(b) along $d$-axis and especially a region bounded by $f$ and an upper dissipation $d_c$ mostly independent of $w$. Here, contrary to the common belief, dissipation anomalously \textit{enhances} CT and reaches fully beyond the possibly unrealistic perturbative regime 
since it is well comparable to other parameters. 
Since the amplitude and phase effect in Eq.~\eqref{eq:CT_main} bears a competition along the $d$-axis, $A_0$ follows below $d_c$ the effect of weakening topological suppression and thus results in this phenomenon universal and robust against modification of parameters. The corresponding instanton pair evolution with $d$ in Fig.~\ref{Fig:Paths}(b) is qualitatively similar to Fig.~\ref{Fig:Paths}(a) in terms of approaching the transition to merged instantons.
Here, we reveal the first realistic scenario of an unconventional dissipative enhancement of the quantum interference tunneling effect. Without any anomalous type of coupling\cite{\mycomment{Leggett1984,*Widom1984,}Leggett1997}, it relies on a generic, although overlooked interplay between the Berry phase and the dissipation-dependent tunneling geometry, which cooperatively contribute to CT.

\section{Dissipative instanton gas}\label{Sec:gas}
Only upon the previous consistent account of single instanton's dissipation and Berry phase for tunneling events, can we further extend to inter-instanton dissipation in an instanton gas with Berry phase effect. \mycomment{with all configurations of an indefinite number of all types of interfering instantons, }The full transition amplitude from $\vec{Q}_-$ to $\vec{Q}_+$ or back to $\vec{Q}_-$ becomes a grand-canonical partition function
$   A(\vec{Q}_-,\vec{Q}_\mp) 
    = \sum_{n=0}^\infty \int_{-T}^ T\dd\tau_1
\int_{-T}^{\tau_1}\dd\tau_2
\cdots
\int_{-T}^{\tau_{n+m-1}}\dd\tau_{n+m} \, y^{n+m} \ee^{-X}$ where $m=n+\frac{1\mp1}{2}$. Logarithmic Coulomb interaction $X(\{\tau_i\}) = -c\sum_{i<j}^{n+m} q_iq_j \ln{|\tau_i-\tau_j|}$ accounts for the inter-instanton dissipation in the dilute approximation with $i$th instanton's charge $q_i=(-1)^{i+1}$. As shown in Append.~\ref{App:gas}, the gas fugacity from instanton tunneling events is calculated to be $y=KA_0$ with $K$ the harmonic fluctuation determinant, where the dissipative Berry phase interference Eq.~\eqref{eq:CT_main} enters. In the noninteracting limit, $A(\vec{Q}_-,\vec{Q}_\mp) \propto \cosh{(2 K TA_0)}$ or $\sinh{(2 K TA_0)}$. Compared with the calculation from the even and odd ground-state subspace, the tunnel splitting $\Delta=KA_0$. It \mycomment{is inapplicable to map this Coulomb gas to the sine-Gordon model\mycomment{\cite{Samuel1978}} due to the instanton time-ordering, while it }formally resembles the anisotropic Kondo problem as per the two-state-flipping nature between $\vec{Q}_\pm$. Renormalization-group analysis\cite{Anderson1971}\mycomment{, this inter-instanton nonlocal interaction} here leads to a critical dissipation $d_0(w)=2 /\mycomment{(2\max{Q_1})^2}{[2Q_1(T)]}^2> 2$. This reassures that our earlier phase diagram region of interest is below the localization transition and the Toulouse limit: coherent or partially damped tunneling mostly remains valid.
Here, we neglect the interaction's dependence on four instanton types, which is in principle possible and otherwise leads to a more complex gas with the fugacity summation inseparable from interaction. 


\section{Conclusion}
We propose and derive a quantum CT model around a conical intersection to address for the first time the hidden and overlooked interplay between dissipation and monopole Berry phase, which has its roots in various mesoscopic processes in solid-state topological materials and chemical reactions. Only by consistently analyzing the tunneling instantons, the associated Berry phases, and the dissipation in a systematic approach, can we reveal a phase diagram with dissipative quantum interference from Berry phase effect. A dissipative Kramers degeneracy, a dissipation-driven phase transition of quantum interference and an unconventional dissipation-enhanced CT emerge as the novel quantum effects. 

\begin{acknowledgments}
X.-X.Z appreciates stimulating discussions with K. Misaki and the Max Planck-UBC-UTokyo Center for Quantum Materials for fruitful collaborations and financial support. This work was supported by NSERC and CIfAR, JSPS KAKENHI (No.~18H03676), and JST CREST (Nos.~JPMJCR16F1 \& JPMJCR1874).
\end{acknowledgments}\mycomment{\Yinyang}

\appendix

\section{Derivation of effective Lagrangian and instanton equation}\label{App:Derivation}
We consider the Hamiltonian with a single spin-\textrm{\textonehalf} or any two-level system coupled to a group of bosonic fields
\begin{equation}\label{eq:H_all}
    H = H_\mathrm{s} + H_\mathrm{b} + H_\mathrm{c} + H_\mathrm{ct}.
\end{equation}
For completeness, we also include the counter term $H_\mathrm{ct} = \sum_{i\nu}{\frac{\tilde{c}_{i\nu}^2}{2m_\nu\omega_\nu^2}\sigma_i\sigma_i}$ with $\tilde{c}_{i\nu}=c_ig_{i\nu}$. $H_\mathrm{ct}$ compensates the side effect of $H_\mathrm{c}$ and retains the original potential surface of $H_\mathrm{s}$, without which $H_\mathrm{c}$ introduces not only dissipation. However, for the two-level nature, $H_\mathrm{ct}$ herein becomes constant and insignificant and is neglected in Sec.~\ref{Sec:model}. 
We work in the imaginary time formalism with $t\rightarrow-\ii\tau$ and hence $\ee^{\ii\mathcal{S}/\hbar}\rightarrow\ee^{-\mathcal{S}/\hbar}$ for any action $\mathcal{S}$ and we use the Fourier convention $\varphi(\tau)=\sum_\omega{\ee^{-\ii\omega\tau}\varphi_\omega}$ and $\varphi_\omega=\frac{1}{\beta}\int_0^\beta{\dd\tau\varphi(\tau)\ee^{\ii\omega\tau}}$ where $\omega$ stands for generic bosonic Matsubara frequencies and we for the nonce set $\tau\in[0,\beta]$ in compliance with the common notation. The corresponding actions are the following ones. 
\begin{equation}
    \mathcal{S}_\mathrm{s} = \mathcal{S}_\mathrm{\Phi} + \int_0^\beta{\dd \tau H_\mathrm{s}}.
\end{equation}
where we include the spin Berry phase term
\begin{equation}\label{eq:spinBerry}
\mathcal{S}_\mathrm{\Phi} = \ii \Phi = \ii S \int_0^\beta{\dd\tau (1-\cos\theta)\dot\phi}   \end{equation}
with polar and azimuthal angle $\theta,\phi$ on the Bloch sphere of the spin $S$.
\begin{widetext}
\begin{equation}
    \mathcal{S}_\mathrm{b} = \int_0^\beta{\dd\tau \sum_{i\nu}{\frac{m_\nu}{2}(\dot{x}_{i\nu}^2+\omega_\nu^2x_{i\nu}^2)}} \mycomment{= \sum_{i\nu}{\frac{m_\nu}{2}\sum_{\omega\omega'}{ [(-\ii\omega)(-\ii\omega')+\omega_\nu^2] x_{\nu\omega}^i x_{\nu\omega'}^i \int_0^\beta{\dd\tau \ee^{-\ii(\omega+\omega')\tau} } } }}
    = \beta\sum_{i\nu}{\frac{m_\nu}{2}\sum_{\omega}{ (\omega^2+\omega_\nu^2) x_{\nu\omega}^i x_{\nu,-\omega}^i } }.
\end{equation}
\begin{equation}
    \mathcal{S}_\mathrm{c} = \int_0^\beta{\dd\tau\sum_i{c_i \sigma_i q_i}} = \beta \sum_{i\omega}{c_i\sigma_{i\omega}q_{i,-\omega}} = \beta \sum_{i\nu\omega}{\tilde{c}_{i\nu}\sigma_{i\omega}x_{i\nu,-\omega}}.
\end{equation}
\begin{equation}
    \mathcal{S}_\mathrm{ct} = \int_0^\beta{\dd\tau\sum_{i\nu}{\frac{\tilde{c}_{i\nu}^2}{2m_\nu\omega_\nu^2}\sigma_i(\tau)\sigma_i(\tau)}} = \beta \sum_{i\nu\omega}{\frac{\tilde{c}_{i\nu}^2}{2m_\nu\omega_\nu^2} \sigma_{i\omega}\sigma_{i,-\omega}}.
\end{equation}
And we introduce the Lagrange multiplier $\lambda_i$ for the collective coordinate $\vec q$
\begin{equation}
    \mathcal{S}_\mathrm{\lambda} = \int_0^\beta{\dd\tau\sum_{i}{\ii\lambda_i(\tau)(q_i(\tau)-\sum_\nu{g_{i\nu}x_{i\nu}})}} = \ii\beta \sum_{i\omega}{\lambda_{i\omega} (q_{i,-\omega}-\sum_\nu{g_{i,\nu}x_{i\nu,-\omega}})} \mycomment{= \ii\beta \sum_{i\omega}{[\lambda_{i\omega} q_{i,-\omega}-\sum_\nu{\frac{g_{i,\nu}}{2}(x_{i\nu\omega}^*\lambda_{i\omega}+\lambda_{i\omega}^*x_{i\nu\omega})}]} }.
\end{equation}
We integrate out all the bosonic bath in the partition function except the collective modes $q_i$
\mycomment{\begin{equation} 
\begin{split}
    \mathcal{Z}[\sigma,x,q,\lambda]=& \int{\mathcal{D}\sigma\mathcal{D}x\mathcal{D}q\mathcal{D}\lambda \, \ee^{ - (\mathcal{S}_\mathrm{s} + \mathcal{S}_\mathrm{b} + \mathcal{S}_\mathrm{c} + \mathcal{S}_\mathrm{ct} + \mathcal{S}_\mathrm{\lambda})}} \\
    =&\int{\mathcal{D}\sigma\mathcal{D}x\mathcal{D}q\mathcal{D}\lambda \, \ee^{ - \left( \mathcal{S}_\mathrm{s} + \mathcal{S}_\mathrm{ct} + 
    \beta\sum_{i\nu\omega}{\{\frac{m_\nu}{2} (\omega^2+\omega_\nu^2) x_{i\nu\omega}^* x_{i\nu\omega} + 
    \frac{1}{2} [ x_{i\nu\omega}^* (\tilde{c}_{i\nu}\sigma_{i\omega} - \ii g_{i\nu}\lambda_{i\omega}) + (\tilde{c}_{i\nu}\sigma_{i\omega}^* - \ii g_{i\nu}\lambda_{i\omega}^*) x_{i\nu\omega} ] \} 
    + \ii\beta \sum_{i\omega}{\lambda_{i\omega} q_{i,-\omega}} }
    \right)}}\\
    =&\mathcal{Z}_{x} \int{\mathcal{D}\sigma\mathcal{D}q\mathcal{D}\lambda \, \ee^{ - \left( \mathcal{S}_\mathrm{s} + \mathcal{S}_\mathrm{ct} + 
     \beta\sum_{i\nu\omega}{ (-B_{\nu\omega}) (\tilde{c}_{i\nu}\sigma_{i\omega} - \ii g_{i\nu}\lambda_{i\omega})  (\tilde{c}_{i\nu}\sigma_{i\omega}^* - \ii g_{i\nu}\lambda_{i\omega}^*)  
    + \ii\beta \sum_{i\omega}{\lambda_{i\omega} q_{i,-\omega}} }
    \right)}}\\
    =&\mathcal{Z}_{x} \int{\mathcal{D}\sigma\mathcal{D}q\mathcal{D}\lambda \, \ee^{ - \left( \mathcal{S}_\mathrm{s} + \mathcal{S}_\mathrm{ct} + 
     \beta\sum_{i\omega}{ \{-w_{i\omega} \sigma_{i\omega}^*\sigma_{i\omega} + u_{i\omega} \lambda_{i\omega}^*\lambda_{i\omega} +
     \ii [ \lambda_{i\omega}^* (\frac{1}{2} q_{i\omega} + v_{i\omega}\sigma_{i\omega}) + (\frac{1}{2} q_{i\omega}^* + v_{i\omega}\sigma_{i\omega}^*) \lambda_{i\omega}  ] \}}
    \right)}}\\
    =&\mathcal{Z}_{x\lambda} \int{\mathcal{D}\sigma\mathcal{D}q \, \ee^{ - \left( \mathcal{S}_\mathrm{s} + \mathcal{S}_\mathrm{ct} + 
     \beta\sum_{i\omega}{ [-w_{i\omega} \sigma_{i\omega}^*\sigma_{i\omega} + u_{i\omega}^{-1} (\frac{1}{2} q_{i\omega} + v_{i\omega}\sigma_{i\omega})  (\frac{1}{2} q_{i\omega}^* + v_{i\omega}\sigma_{i\omega}^*)  ]}
    \right)}}\\
    =&\mathcal{Z}_{x\lambda} \int{\mathcal{D}\sigma\mathcal{D}q \, \ee^{ - \left( \mathcal{S}_\mathrm{s} + \mathcal{S}_\mathrm{ct} + 
     \beta\sum_{i\omega}{ [ u_{i\omega}^{-1} q_{i\omega}^* q_{i\omega} /4 + c_i(\sigma_{i\omega}^*q_{i\omega}+q_{i\omega}^*\sigma_{i\omega})/2  ]}
    \right)}}\\
    =&\mathcal{Z}_{x\lambda q} \int{\mathcal{D}\sigma\mathcal{D}q \, \ee^{ - \left( \mathcal{S}_\mathrm{s} + \mathcal{S}_\mathrm{ct} + 
     \beta\sum_{i\omega}{ (  -c_i^2u_{i\omega}\sigma_{i\omega}^*\sigma_{i\omega}  )}
    \right)}}\\
\end{split}
\end{equation}}
\begin{equation} \label{eq:gaussianInt}
\begin{split}
    \mathcal{Z}[\sigma,x,q,\lambda]=& \int{\mathcal{D}\sigma\mathcal{D}x\mathcal{D}q\mathcal{D}\lambda \, \ee^{ - (\mathcal{S}_\mathrm{s} + \mathcal{S}_\mathrm{b} + \mathcal{S}_\mathrm{c} + \mathcal{S}_\mathrm{ct} + \mathcal{S}_\mathrm{\lambda})}} \\
    =&\int{\mathcal{D}\sigma\mathcal{D}x\mathcal{D}q\mathcal{D}\lambda \, \ee^{ - \left( \mathcal{S}_\mathrm{s} + \mathcal{S}_\mathrm{ct} + 
    \beta\sum_{i\nu\omega}{\{\frac{m_\nu}{2} (\omega^2+\omega_\nu^2) x_{i\nu\omega}^* x_{i\nu\omega} + 
     (\tilde{c}_{i\nu}\sigma_{i\omega}^* - \ii g_{i\nu}\lambda_{i\omega}^*) x_{i\nu\omega}  \} 
    + \ii\beta \sum_{i\omega}{\lambda_{i\omega} q_{i,-\omega}} }
    \right)}}\\
    =&\mathcal{Z}_{x} \int{\mathcal{D}\sigma\mathcal{D}q\mathcal{D}\lambda \, \ee^{ - \left( \mathcal{S}_\mathrm{s} + \mathcal{S}_\mathrm{ct} + 
     \beta\sum_{i\nu\omega}{ (-B_{\nu\omega})  (\tilde{c}_{i\nu}\sigma_{i\omega}^* - \ii g_{i\nu}\lambda_{i\omega}^*)  (\tilde{c}_{i\nu}\sigma_{i\omega} - \ii g_{i\nu}\lambda_{i\omega}) 
    + \ii\beta \sum_{i\omega}{\lambda_{i\omega} q_{i,-\omega}} }
    \right)}}\\
    =&\mathcal{Z}_{x} \int{\mathcal{D}\sigma\mathcal{D}q\mathcal{D}\lambda \, \ee^{ - \left( \mathcal{S}_\mathrm{s} + \mathcal{S}_\mathrm{ct} + 
     \beta\sum_{i\omega}{ \{-w_{i\omega} \sigma_{i\omega}^*\sigma_{i\omega} + \frac{1}{4} u_{i\omega} \lambda_{i\omega}^*\lambda_{i\omega} +
     \ii [ \lambda_{i\omega}^* (\frac{1}{2} q_{i\omega} + v_{i\omega}\sigma_{i\omega}) + (\frac{1}{2} q_{i\omega}^* + v_{i\omega}\sigma_{i\omega}^*) \lambda_{i\omega}  ] \}}
    \right)}}\\
    =&\mathcal{Z}_{x\lambda} \int{\mathcal{D}\sigma\mathcal{D}q \, \ee^{ - \left( \mathcal{S}_\mathrm{s} + \mathcal{S}_\mathrm{ct} + 
     \beta\sum_{i\omega}{ [-w_{i\omega} \sigma_{i\omega}^*\sigma_{i\omega} + 4u_{i\omega}^{-1} (\frac{1}{2} q_{i\omega} + v_{i\omega}\sigma_{i\omega})  (\frac{1}{2} q_{i\omega}^* + v_{i\omega}\sigma_{i\omega}^*)  ]}
    \right)}}\\
    =&\mathcal{Z}_{x\lambda} \int{\mathcal{D}\sigma\mathcal{D}q \, \ee^{ - \left( \mathcal{S}_\mathrm{s} + \mathcal{S}_\mathrm{ct} + 
     \beta\sum_{i\omega}{ [ u_{i\omega}^{-1} q_{i\omega}^* q_{i\omega}  + c_i(\sigma_{i\omega}^*q_{i\omega}+q_{i\omega}^*\sigma_{i\omega})/2  ]}
    \right)}}\\
    =&\mathcal{Z}_{x\lambda q} \int{\mathcal{D}\sigma \, \ee^{ - \left( \mathcal{S}_\mathrm{s} + \mathcal{S}_\mathrm{ct} + 
     \beta\sum_{i\omega}{ (  -w_{i\omega}\sigma_{i\omega}^*\sigma_{i\omega}  )}
    \right)}}\\
\end{split}
\end{equation}
where we introduce for the sake of notational brevity $A_i=\sum_\nu{\frac{\tilde{c}_{i\nu}^2}{2m_\nu\omega_\nu^2}}$, $B_{\nu\omega}=\frac{1}{2m_\nu(\omega^2+\omega_\nu^2)}$, $u_{i\omega}=4\sum_\nu{B_{\nu\omega}g_{i\nu}^2}$, $w_{i\omega}=\sum_\nu{B_{\nu\omega}\tilde{c}_{i\nu}^2}=c_i^2u_{i\omega}/4$, $v_{i\omega}=\sum_\nu{B_{\nu\omega}g_{i\nu}\tilde{c}_{i\nu}}=c_iu_{i\omega}/4$ and absorb the generated determinants into the prefactors.
From the penultimate line of Eq.~\eqref{eq:gaussianInt}, we obtain the effective action $\mathcal{S}_\mathrm{eff}$ for the spin system coupled to the collective mode $\vec{q}$
\begin{equation}\label{eq:S_eff}
    \mathcal{S}_\mathrm{eff} = \mathcal{S}_\mathrm{s} +  \beta\sum_{i\omega}{ [ u_{i\omega}^{-1} q_{i\omega}^* q_{i\omega} + c_i\sigma_{i\omega}^*q_{i\omega}  + A_i\sigma_{i\omega}^*\sigma_{i\omega}]} 
    \mycomment{&= \mathcal{S}_\mathrm{s} + \sum_{i}{ [ \int_0^\beta {\dd\tau\dd\tau' \frac{1}{\beta}u_{i}^{-1}(\tau-\tau') q_{i}(\tau) q_{i}(\tau')} + \int_0^\beta {\dd\tau(c_i\sigma_{i}(\tau)q_{i}(\tau)  + A_i\sigma_{i}(\tau)\sigma_{i}(\tau))}]} \\ }
    = \mathcal{S}_\mathrm{s} +  \mathcal{S}_\mathrm{D} + \mathcal{S}_\mathrm{c} + \mathcal{S}_\mathrm{ct}
\end{equation}
where 
\begin{equation}\label{eq:S_D}
    \mathcal{S}_\mathrm{D} = \frac{1}{\beta}\sum_{i}{  \int_0^\beta {\dd\tau\dd\tau' u_{i}^{-1}(\tau-\tau') q_{i}(\tau) q_{i}(\tau')}}    
\end{equation}
is the newly generated temporally nonlocal interaction responsible for dissipation.
And also we have the pure spin effective action $\mathcal{S}_\mathrm{eff}'$ from the last line of Eq.~\eqref{eq:gaussianInt} where the counter term $\mathcal{S}_\mathrm{ct}$ cancels partially the action
\begin{equation}\label{eq:S_eff2}
    \mathcal{S}_\mathrm{eff}' = \mathcal{S}_\mathrm{s} + \beta\sum_{i\nu\omega}{  \frac{\tilde{c}_{i\nu}^2}{2m_\nu} \frac{\omega^2}{\omega_\nu^2(\omega^2+\omega_\nu^2)} \sigma_{i\omega}^*\sigma_{i\omega} } 
    =  \mathcal{S}_\mathrm{s} + \mathcal{S}_\mathrm{i},
\end{equation}
where 
\begin{equation}\label{eq:S_i}
    \mathcal{S}_\mathrm{i}=\frac{1}{\beta} \int_0^\beta {\dd\tau\dd\tau' \sum_{i}{ \sigma_{i}(\tau)\sigma_{i}(\tau') K_i(\tau-\tau') }  }
\end{equation}
is the interaction term nonlocal in time,
\begin{equation}\label{eq:K_i}
    K_i(\tau)=\int_0^\infty{ \frac{\dd\omega'}{\pi} J_i(\omega') D_{\omega'}(\tau) }    
\end{equation}
is the temporally nonlocal kernel,
\begin{equation}\label{eq:J_i}
    J_i(\omega) = \frac{\pi}{2} \sum_\nu{ \frac{\tilde{c}_{i\nu}^2}{m_\nu\omega_\nu} \delta(\omega-\omega_\nu) }  
\end{equation}
is the spectral density of the coupling to the bosonic environment and $D_{\omega'}(\tau) = \sum_\omega{ \frac{\omega^2}{\omega'(\omega^2+\omega'^2)} \ee^{\ii\omega\tau}}$.

In reality, the spectral density $J_i$ in Eq.~\eqref{eq:J_i} must differ from the ideal memoryless friction of $J_0(\omega) = \eta\omega$ and decay fast enough as $\omega\rightarrow\infty$. Here we adopt an analytically tractable Lorentzian-like regularization with cutoff frequency $\omega_\mathrm{D}$ 
\begin{equation}\label{eq:cutoff}
    J_i(\omega) = \eta_i\omega^s/(1+\omega^2/\omega_\mathrm{D}^2)^2
\end{equation}
and the corresponding memory-friction kernel $\gamma(\omega)$, also known as the damping function in the generalized Langevin equation, will acquire a memory-friction time scale $1/\omega_\mathrm{D}$. The low-frequency behavior does not depend on the specific regularization form as long as the decay is fast enough.
Then $A_i=\frac{1}{\pi} \int{ \frac{\dd\omega}{\omega} J_i(\omega) }\mycomment{ = \red{???}}$ and when $s<4$
\begin{equation}
    u_{i\omega} = \frac{4}{\pi c_i^2} \int{ \dd\omega' J_i(\omega') \frac{\omega'}{\omega^2+\omega'^2} } 
    = \frac{\eta_i}{c_i^2} \frac{\omega_\mathrm{D}^2 [ \omega_\mathrm{D}^s (\omega^2s+(2-s)\omega_\mathrm{D}^2) - 2\omega_\mathrm{D}^2 |\omega^s| ]}{ (\omega^2-\omega_\mathrm{D}^2)^2} \csc{\frac{\pi s}{2}}.
\end{equation}
For the ohmic dissipation $s=1$, we have $u_{i\omega} = \frac{\eta_i}{\pi c_i^2} \frac{\pi \omega_\mathrm{D}^3}{(\omega_\mathrm{D}+|\omega|)^2}$ and $A_i=\eta_i\omega_\mathrm{D}/4$. Henceforth, we will set $\eta_i=\eta$ for simplicity and $\mathcal{S}_\mathrm{ct}$ becomes an insignificant constant in the coherent state spin path integral.
In the following, let's study the kernel $X_{i\omega} = (u_{i\omega})^{-1} = \frac{\kappa_i}{\omega_\mathrm{D}^2} (\omega_\mathrm{D}^2+2\omega_\mathrm{D}|\omega|+\omega^2)$ that appears in Eq.~\eqref{eq:S_eff} and we denote $\kappa_i=\frac{c_i^2}{\eta_i\omega_\mathrm{D}}$. In the imaginary-time domain, we have
\begin{equation}
    X_i(\tau,\tau') = \beta\kappa_i\delta(\tau-\tau') + K_i^*(\tau-\tau') + \frac{\kappa_i}{\omega_\mathrm{D}^2} \partial_\tau\partial_{\tau'}\delta(\tau-\tau')
\end{equation}
where $K_i^*(\tau)=\frac{2\kappa_i}{\omega_\mathrm{D}}\mathcal{F}[|\omega|]$ and $\mathcal{F}$ means the Fourier transform. We digress a little on the evaluation of $K_i^*$ and consider a general $k(\tau)=\frac{\xi}{2}\mathcal{F}[|\omega|]$. We express the Fourier component of $k(\tau)$, $k_n = \frac{\xi}{2} |\omega_n|$, by introducing a spectral density $J(\omega)=\xi\omega$ and follow the summation form in Eqs.~\eqref{eq:K_i}\eqref{eq:J_i}
\begin{equation}
     k_n = \sum_\nu{ \frac{c_\nu^2}{2m_\nu\omega_\nu^2} \frac{\omega_n^2}{\omega_\nu^2+\omega_n^2} } = \frac{\xi}{2} |\omega_n| = \frac{\omega_n^2}{\pi} \int_0^\infty{ \dd\omega \frac{J(\omega)}{\omega(\omega^2+\omega_n^2)} }
\end{equation}
where we restore the Matsubara index $n$ for clearness. Then we have the following properties:
1) Because of $k_n=k_{-n}$ we have $k(\tau) = k(\beta-\tau)$;
2) The $0$th Fourier component $\frac{1}{\beta} \int_0^\beta{ \dd\tau k(\tau) } = k_{n=0} = 0$;
3) We can define $K(\tau) = \beta\zeta \delta(\tau)\mycomment{:\!\delta(\tau)\!:} - k(\tau)$ where $\zeta = \sum_\nu{ \frac{c_\nu^2}{2m_\nu\omega_\nu^2} = \frac{1}{\pi} \int_0^\beta{\dd\omega \frac{J(\omega)}{\omega}} }$\mycomment{ and $:\delta(\tau): = \sum_n{\ee^{\ii\omega_n\tau}}/\beta =  \sum_n{\delta(\tau-n\beta)}$ is the periodically continued $\delta$-function};
4) $K_n = \zeta - k_n = \sum_\nu{ \frac{c_\nu^2}{2m_\nu(\omega_\nu^2+\omega_n^2)}  
    = \frac{1}{\pi} \int_0^\beta{\dd\omega \frac{J(\omega)\omega}{\omega^2+\omega_n^2}} }$.
We can thus obtain 
\begin{equation}
\begin{split}
    \beta\sum_{\omega_n}{ k_n q_n^* q_n } &= \frac{1}{\beta} \int_0^\beta {\dd\tau\dd\tau' k(\tau-\tau') q_{i}(\tau) q_{i}(\tau')}
    \LongEqual{1)} \frac{2}{\beta} \int_0^\beta {\dd\tau \int_0^\tau {\dd\tau' k(\tau-\tau') q_{i}(\tau) q_{i}(\tau')}} \\
    &\LongEqual{2)} -\frac{1}{\beta} \int_0^\beta {\dd\tau \int_0^\tau {\dd\tau' k(\tau-\tau') (q_{i}(\tau) - q_{i}(\tau'))^2}} 
    \LongEqual{3)} \frac{1}{\beta} \int_0^\beta {\dd\tau \int_0^\tau {\dd\tau' K(\tau-\tau') (q_{i}(\tau) - q_{i}(\tau'))^2}}.
\end{split}
\end{equation}
\begin{equation}
    K(\tau) = \sum_n{ K_n \ee^{\ii\omega_n\tau} }
    \LongEqual{4)} \frac{\beta}{2\pi} \int_0^\beta{\dd\omega J(\omega)\frac{\cosh{[\omega(\beta/2-\tau)]}}{\sinh{(\omega\beta/2)}}}
    = \xi \frac{\beta}{2\pi} \frac{(\pi/\beta)^2}{\sin^2{(\pi\tau/\beta)}}.
\end{equation}
Therefore, $\mathcal{F}[|\omega|] = \frac{2}{\xi}\beta\zeta\delta(\tau) - \frac{\beta}{\pi} \frac{(\pi/\beta)^2}{\sin^2{(\pi\tau/\beta)}}$\mycomment{ and $K_i^*(\tau) = \frac{2\kappa_i}{\omega_\mathrm{D}} \frac{(\pi/\beta)}{\sin^2{(\pi\tau/\beta)}}$}.
When $\tau\ll\beta$ as required by our focus of the zero-temperature case, we can make the replacement 
$\frac{\pi}{\beta^2} \frac{1}{\sin^2{[\pi(\tau-\tau')/\beta]}} 
\rightarrow \frac{1}{\pi} \frac{1}{(\tau-\tau')^2}$.
And our effective action becomes
\begin{equation}\label{eq:S_final}
    \mathcal{S} = \mathcal{S}_\mathrm{\Phi}+ \int_0^\beta{ \dd\tau [\sum_i \frac{1}{2} m_i\dot{q}_i^2(\tau) + V(\vec q) +  \sum_i \int_0^\beta{ \dd\tau' \frac{d_i}{4} \frac{(q_i(\tau)-q_i(\tau'))^2}{(\tau-\tau')^2}  } ] }.
\end{equation}
where $m_i = \frac{2\kappa_i}{\omega_\mathrm{D}^2}, d_i = \frac{4\kappa_i}{\pi\omega_\mathrm{D}}, \mu_i=\kappa_i, V(\vec q) = -|\vec Q| + \sum_i{\mu_iq_i^2(\tau)}$. Here $m_i,d_i,\mu_i$ apparently have two degrees of freedom. It is important to note that this is merely an artifact of the simple form of regularization in Eq.~\eqref{eq:cutoff}. In general, those three parameters are independent and are the leading terms generated from integrating out the environment. Henceforth, we will treat them as free parameters. 

For the sake of later instanton solutions under general BCs, we note that $\beta$ should be replaced by an imaginary time $2T$ not necessarily related to temperature, since we are actually calculating the transition amplitude from the initial position $\vec{Q}^\mathrm{i}$ to the final position $\vec{Q}^\mathrm{f}$ in imaginary time $[-T,T]$
\begin{equation}
    \langle \vec{Q}^\mathrm{f}\lvert \ee^{-2\mathcal{H}T} \rvert\vec{Q}^\mathrm{i}\rangle=\int\limits_{\vec{Q}(-T)=\vec{Q}^\mathrm{i}}^{\vec{Q}(T)=\vec{Q}^\mathrm{f}} \mathcal{D} \vec{Q}(\tau) \ee^{-\mathcal{S}[\vec{Q}(\tau)]},
\end{equation}
which takes the form of a partition function.
Let us now change the variable to $\vec Q$ and nondimensionalize $\mathcal{S}_\mathrm{Q}$ in the total action Eq.~\eqref{eq:S_final} $\mathcal{S} = \mathcal{S}_\mathrm{\Phi} + \mathcal{S}_\mathrm{Q}$
\begin{equation}\label{eq:S_eff3}
\begin{split}
\mathcal{S}_\mathrm{Q} &=  \int_{-T}^{T}{ \dd\tau [\sum_i\frac{m_i}{2c_i^2}\dot{Q}_i^2 + V(\vec Q) ] 
+ \int_{-T}^{T}\int_{-T}^{T}{ \dd\tau\dd\tau' \sum_i\frac{d_i}{4 c_i^2} \frac{(Q_{i\tau}-Q_{i\tau'})^2}{(\tau-\tau')^2}  }  } \\
\mycomment{&= \mathcal{S}_\mathrm{\Phi}+  \int_0^{\tilde\beta}{ \tau_0\dd x [\sum_i\frac{1}{\tau_0^2} \frac{m_i}{2c_i^2}\left(\frac{\dd{Q}_i}{\dd x}\right)^2 + V(\vec Q) ] 
+ \int_0^{\tilde\beta}\int_0^{\tilde\beta}{ \tau_0^2\dd x\dd x' \sum_i\frac{d_i}{4\pi c_i^2} \frac{1}{\tau_0^2} \frac{(Q_{ix}-Q_{ix'})^2}{(x-x')^2}  }  } \\}
&=  \int_{-\tilde T}^{\tilde T}{ \dd \tilde{\tau} [ \sum_i \frac{m_i}{2c_i^2\tau_0}\left(\frac{\dd{Q}_i}{\dd \tilde{\tau}}\right)^2 + \tau_0V(\vec Q) ] 
+ \int_{-\tilde T}^{\tilde T}\int_{-\tilde T}^{\tilde T}{ \dd \tilde{\tau}\dd \tilde{\tau}' \sum_i\frac{d_i}{4 c_i^2}  \frac{(Q_{i\tilde{\tau}}-Q_{i\tilde{\tau}'})^2}{(\tilde{\tau}-\tilde{\tau}')^2}  }  } \\
\mycomment{&= \mathcal{S}_\mathrm{\Phi} + \alpha_0\tau_0 \left( \int_{-\tilde T}^{\tilde T}{ \dd \tilde{\tau} [ \sum_i\frac{\tilde{m}_i}{2\tilde{c}_i^2}\left(\frac{\dd\tilde{Q}_i}{\dd \tilde{\tau}}\right)^2 + \tilde{V}(\vec{\tilde{Q}}) ] }
+ \int_{-\tilde T}^{\tilde T}\int_{-\tilde T}^{\tilde T}{ \dd \tilde\tau\dd \tilde\tau' \sum_i \frac{\tilde{d}_i}{4\tilde{c}_i^{2}} \frac{(\tilde{Q}_{i\tilde\tau}-\tilde{Q}_{i\tilde\tau'})^2}{(\tilde\tau-\tilde\tau')^2}  }  \right) \\}
&=  \alpha_0\tau_0 \left( \int_{-\tilde T}^{\tilde T}{ \dd \tilde{\tau} [ \sum_i\frac{\tilde{m}_i}{2}\left(\frac{\dd\tilde{Q}_i}{\dd \tilde{\tau}}\right)^2 + \tilde{V}(\vec{\tilde{Q}}) ] }
+ \int_{-\tilde T}^{\tilde T}\int_{-\tilde T}^{\tilde T}{ \dd \tilde{\tau}\dd \tilde{\tau}' \sum_i \frac{\tilde{d}_i}{4} \frac{(\tilde{Q}_{i\tilde{\tau}}-\tilde{Q}_{i\tilde\tau'})^2}{(\tilde\tau-\tilde\tau')^2}  }  \right)
\end{split}
\end{equation}
where
\begin{equation}\label{eq:potential}
    \tilde{V}(\vec{\tilde{Q}}) =  \sum_i{ (\tilde{Q}_i - \tilde{w}_i)^2 / \tilde{\alpha}_i - |\vec{\tilde{Q}}| }
\end{equation}
and we introduce $\alpha_i=c_i^2/\mu_i$. All dimensional quantities are nondimensionalized as the following  $\tilde{\alpha}_i=\alpha_i/\alpha_0$, $\tilde{w}_i=w_i/\alpha_0$, $\tilde{Q}_i=Q_i/\alpha_0$, \mycomment{$\tilde{m}_i=m_i/m_0$}$\tilde{m}_i=\frac{m_i}{m_0\tilde{c}_i^2}$, $\tilde{\tau}=\tau/\tau_0$, $\tilde T = T/\tau_0$, $\tilde{c}_i=c_i/\sqrt{\alpha_0\mu_0}$, $\tilde{d}_i=\frac{d_i}{\tilde{c}_i^2\sqrt{m_0\mu_0}}$ with $\tau_0=\sqrt{m_0/\mu_0}$. Therefore, one needs to choose three dimensionful base parameters $\alpha_0,m_0,\mu_0$ in total. As per the adiabatic approximation and the original realness of the coordinate $\vec{Q}$, we proceed by solving the path for the real-valued $\mathcal{S}_\mathrm{Q}$ while the imaginary $\mathcal{S}_\mathrm{\Phi}$ solely attaches a complex phase to the quantum amplitude. The instanton Euler-Lagrange semiclassical equation of motion (EOM) $\frac{\delta \mathcal{S}_\mathrm{Q}}{\delta Q_i}=0$ takes the form 
\begin{equation}\label{eq:EOM}
\tilde{m}_i\frac{\dd^2\tilde{Q}_i}{{\dd \tilde{\tau}}^2} - \frac{\partial\tilde{V}(\vec{\tilde{Q}})}{\partial \tilde{Q}_i} 
- \tilde{d}_i \int_{-\tilde T}^{\tilde T}{ \dd \tilde{\tau}' \frac{(\tilde{Q}_{i\tilde{\tau}}-\tilde{Q}_{i\tilde{\tau}'})}{(\tilde\tau-\tilde{\tau}')^2} }  = 0,
\end{equation}
in which $\tilde{m}_i,\tilde{d}_i$ reduces to $M_i,D_i$ given in Sec.~\ref{Sec:model} when we impose the isotropic parameter choices $m_i=m_0,d_i=d\sqrt{m_0\mu_0},\mu_i=\mu_0$. Elsewhere in the paper, we drop the tilde symbol for brevity.  Note that this form is in the transformed monopole-centered $\vec{Q}$ coordinate system. Without much loss of generality and to simplify the discussion henceforth, we will set in the original coordinate $\vec{q}$ all parameters, including the mass $\vec{m}$, the dissipation strength $\vec{d}$, and the generated harmonic potential $\vec{\mu}$, isotropic but the spin Zeeman-like field $\vec{w}$ and the effective coupling $\vec{\alpha}$. It describes effectively in the $\vec{q}$-space ($\vec{Q}$-space) that the instanton quasiparticle with (an)isotropic mass and dissipation moves under an anisotropic potential $\tilde{V}$, where $\tilde{m}_i,\tilde{d}_i\propto\tilde\alpha_i^{-1}$. Effectively, the $\vec{Q}$-space EOMs can be seen in each direction to have isotropic mass and dissipation but a force $\tilde\alpha_i\partial\tilde{V}/\partial \tilde{Q}_i$.
Another choice would be setting $\vec{\tilde{m}},\vec{\tilde{d}}$ isotropic directly in the $\vec{Q}$ space, which corresponds to the case where all parameters are isotropic but $\vec w$ and $\vec \mu$. However, the instanton EOMs are still of the same type with isotropic mass and dissipation and a force $\partial\tilde{V}/\partial \tilde{Q}_i$ instead and only insignificant quantitative difference can occur. All analytic and numerical discussions in this work are confirmed to hold in general. 

\end{widetext}
\section{Instanton potential landscape}\label{App:landscape}
Let us now inspect the potential landscape $V(\vec{Q})$. 
Its extremum manifold is given by $\frac{\partial V(\vec{Q})}{\partial Q_i} = \frac{2}{\alpha_i}(Q_i-w_i) - \frac{Q_i}{|\vec{Q}|} = 0$. We have all $\alpha_i>0$. To determine the nature of these extrema, we calculate the three eigenvalues $\vec \lambda = (\lambda_1,\lambda_2,\lambda_3)$ or the principal minors of the Hessian matrix $\frac{\partial^2 V}{\partial Q_i \partial Q_j}$. Local minima are assured if all eigenvalues or all principal minors are positive.

Let us scan the whole positive octant in the 3D $\vec w$-space and assume $\alpha_1>\alpha_{2,3}$ without loss of generality.
\begin{enumerate}
    \item $\vec w= w_2 \hat 2 + w_3 \hat 3$ (CT)\\
    Two degenerate local minima $V_\pm=-(\frac{\alpha_1}{4}+\sum_{i=2,3}{\frac{w_i^2}{\alpha_1-\alpha_i}})$ at $\vec Q_\pm = (\pm\alpha_1\sqrt{\frac{1}{4}-\sum_{i=2,3}{\frac{w_i^2}{(\alpha_1-\alpha_i)^2}}},\frac{\alpha_1w_2}{\alpha_1-\alpha_2},\frac{\alpha_1w_3}{\alpha_1-\alpha_3})$ that lie on the sphere $|\vec{Q}|=\frac{\alpha_1}{2}$. Using the principal minors of the Hessian, the necessary and sufficient condition for two local minima is $\frac{1}{4}-\sum_{i=2,3}{\frac{w_i^2}{(\alpha_1-\alpha_i)^2}}>0$. A simpler sufficient condition is $\alpha_1-\max[\alpha_2,\alpha_3]>2|\vec w|$. This corresponds to the CT case referred in Sec.~\ref{Sec:EOM}.
    \item $\vec w= w_1 \hat 1 + w_3 \hat 3$ (QD)\\
    Now we have $\frac{2Q_2}{\alpha_2} = \frac{Q_2}{|\vec Q|}$, which leads to $|\vec{Q}|=\frac{\alpha_2}{2}$ or $Q_2=0$. Note that the former, same as the previous case up to an exchange of the coordinates $1$ and $2$, does not give two local minima under our assumption $\alpha_1>\alpha_{2,3}$. Therefore, the only possibility left, the latter case with $Q_2=0$, gives two nondegenerate local minima at $\vec{Q}_\pm=( w_1+\frac{\alpha_1}{2}\cos\theta_\pm, 0, w_3+\frac{\alpha_3}{2}\sin\theta_\pm )$ where $\theta_\pm$ takes two solutions from $\frac{\cos\theta}{\sin\theta}=\frac{w_1+\frac{\alpha_1}{2}\cos\theta}{w_3+\frac{\alpha_3}{2}\sin\theta}$, which amounts to a quartic equation with no concise form of roots. A simple sufficient condition is $\alpha_1-\alpha_3>4|\vec{w}|,\alpha_1-\alpha_2>2|\vec{w}|$.
    \item $\vec w= w_1 \hat 1 + w_2 \hat 2$ (QD)\\
    Same as the previous case up to an exchange of the coordinates $2$ and $3$.
    \item $\vec w= w_1 \hat 1 + w_2 \hat 2 + w_3 \hat 3$ (QD)\\
    There is no simple analytic form of the position $\vec{Q}_\pm$ of the two nondegenerate local minima. A simple sufficient condition is $\alpha_1-\max[\alpha_2,\alpha_3]>4|\vec{w}|$.
\end{enumerate}
Next, we look into the case with $\vec w$ aligned with one axis. Without loss of generality, we assume $\vec{w}=w_3\hat{3}$ but do not put any restriction on $\vec{\alpha}$. There are three mutually exclusive cases.
\begin{enumerate}
\item $\alpha_1=\alpha_2$ continuous manifold of extrema
\begin{enumerate}
    \item $\alpha_1=\alpha_2=\alpha_3=\alpha$\\
  The solution is an $S^2$-sphere given by $|\vec Q|=\alpha/2$.
  \item $\alpha_1=\alpha_2=\alpha_{12}\neq\alpha_3$\\
  The solution is an $S^1$-ring given by $|\vec Q|=\alpha_{12}/2,Q_3=\frac{\alpha_{12}w_3}{\alpha_{12}-\alpha_3}$.
\end{enumerate}
  \item $\alpha_1\neq\alpha_2$
  \begin{enumerate}
     \item $\alpha_1\neq\alpha_3,Q_1\neq0$ (CT)\\
     The solution consists of two points $\vec Q_\pm = (\pm\alpha_1\sqrt{\frac{1}{4}-(\frac{w_3}{\alpha_1-\alpha_3})^2}, 0, \frac{\alpha_{1}w_3}{\alpha_{1}-\alpha_3})$ on the ring specified by $|\vec{Q}|=\alpha_1/2,Q_2=0$ with $\vec{\lambda}=(\frac{2}{\alpha_2}-\frac{2}{\alpha_1}, \frac{1}{\alpha_1\alpha_3}(\alpha_1-\Delta), 
    \frac{1}{\alpha_1\alpha_3}(\alpha_1+\Delta)>0$ where $\Delta=\sqrt{\frac{(\alpha_1-2\alpha_3)^2(\alpha_1-\alpha_3)+16\alpha_3w_3^2}{\alpha_1-\alpha_3}}$. Under the necessary and sufficient condition $\alpha_1>\alpha_{2,3},\alpha_1-\alpha_3>2w_3$, we have two degenerate minima $V_\pm=-(\frac{\alpha_1}{4}+\frac{w_3^2}{\alpha_1-\alpha_3})$ at $\vec Q_\pm$ that lie symmetrically in the upper half of the $Q_1Q_3$-plane. We have a type-II Weyl crossing point\cite{Soluyanov2015} when $\alpha_3<2w_3$.  This matches the general discussion for $\vec w= w_2 \hat 2 + w_3 \hat 3$ and corresponds to the CT case referred in Sec.~\ref{Sec:EOM}.
     \item $\alpha_2\neq\alpha_3,Q_1=0$ (CT)\\
     Same as the previous case up to an exchange of coordinate $1,2$.
  \end{enumerate}
  \item (QD) For general $\vec \alpha$, there is yet another solution $\vec{Q}_\pm = (0,0,w_3\pm\frac{\alpha_3}{2})$. $\vec \lambda = (\frac{2}{\alpha_1}-\frac{2}{|\alpha_3\pm2w_3|}, \frac{2}{\alpha_2}-\frac{2}{|\alpha_3\pm2w_3|}, \frac{2}{\alpha_3})>0$. Under the necessary and sufficient condition $\alpha_3-2w_3>\max[\alpha_1,\alpha_2]>0$, we have two nondegenerate local minima $V_\pm=-|w_3\pm\frac{\alpha_3}{2}|+\frac{\alpha_3}{4}$ at $\vec{Q}_\pm$ and the conical crossing $V_0=\frac{w_3^2}{\alpha_3}>V_->V_+$ at $|\vec{Q}|=0$. If we instead set $\vec{w}=w_1\hat{1}$ that means an exchange of the coordinates $1$ and $3$, it matches the general discussion for $\vec w= w_1 \hat 1 + w_3 \hat 3$ when $\cos\theta=0$ and corresponds to the QD case referred in Sec.~\ref{Sec:EOM}.
\end{enumerate}

\section{Numerical method}\label{App:FDM}
We provide some comments and the details of our approach to the BVP of a system of nonlinear integro-differential equations with a Fredholm integral.
Firstly, any BVP is fundamentally distinct from the initial value problem (IVP). An IVP of classical particle scattering has been discussed in the presence of monopolar field\cite{Misaki2018}. 
Secondly, this BVP cannot be converted into an ordinary differential equation (ODE) problem by differentiating the system and linear integral transforms like the Laplace transform are of no use. Thirdly, there are in principle other methods one may resort to. In an iterative method one replaces the unknown function in the integrand by the solution from the previous round or by an ansatz or guess when it is the first round, and then solves the resulting ODE repeatedly until convergence. Besides, one can also conceptually approximate the solution by a polynomial of a certain degree and thus justify differentiating the system enough times and dropping the Fredholm integral, then followed by shooting methods for BVP for instance. Other possibilities may include differential transform method\cite{Arikoglu2005}, Chebyshev decomposition method\cite{Trefethen2017}, and various asymptotic techniques\cite{He2006}. However, in the methods above, the procedure is sometimes not well-controlled and may heavily depend on the initial guess due to nonlinearity: hardship in achieving correct convergence significantly augments the computational cost.

Here, we solve this problem efficiently combining the finite difference method (FDM)\cite{Leveque2007,Trefethen2017} and the Gaussian quadrature rule\cite{AbramowitzStegun}. 
\begin{itemize}
    \item The Gaussian quadrature rule with $n$ points helps evaluate the integral as a weighted sum based on a class of orthogonal polynomials and is accurate for polynomial integrands of degree $2n-1$ or less. Here, as the integrand is free from endpoint singularities and integrated over a finite region, we adopt the Gauss-Legendre quadrature using Legendre polynomials. We use finite but large enough range of imaginary time $[-T,T]$ to assure convergence of the solution. The quadrature grid nodes are the roots $x_i$ of the Legendre polynomial $P_n(x)$ and the weights are given by $w_i=\frac{2}{(1-x_i^2)P_n'(x_i)}$. The change of integration interval is given by $\int_a^b f(x)\dd x = \frac{b-a}{2} \int_{-1}^1 f\left(\frac{b-a}{2}\xi + \frac{a+b}{2}\right)\dd\xi$. The grid nodes and weights can be generated using the Golub-Welsch algorithm\cite{Golub1969} or the Laurie's algorithm\cite{Laurie1999} based on the positive-definite real symmetric Jacobi matrix. Both the differentiation and integration are processed as per the grid from the Gauss-Legendre quadrature as the integral requires more careful and stringent discretization than the equally spaced Newton-Cotes rules while the differentiation is less susceptible. 
    \item We do not utilize adaptive quadratures with refined subdivisions as we intend to solve the integro-differential equation system at once in an FDM manner. This, however, requires a generic generation of finite difference formulae on arbitrary grids, i.e., nonuniform or irregular stencils. Generic finite difference formulae are equivalent to derivatives of Lagrange interpolating polynomials, for which Taylor expansion is exact. In general, a finite difference formula of $t$-th order derivative on $s$-point data has at least asymptotic order $s-t$ for the error reduction and is exact for polynomials of degree $s-1$. The approximation is better on uniform grids with centered differences. Here, we can use Fornberg's efficient recursion algorithm for finite difference weights on irregular stencils based on polynomial interpolation\cite{Fornberg1988\mycomment{,Fornberg1992,Fornberg1998}}. We apply (partially) one-sided forward or backward formulae near the edges and centered formulae in the bulk. As our problem bears no Gibb's oscillation, a good balance between numerical roundoff error and systematic approximation error is achieved using fourth-order differences, i.e., asymptotic order of error reduction $O(h^4)$.
    \item Consequently, the system of three nonlinear integro-differential equations is converted to a system of $3m$ nonlinear algebraic equations of $3m$ unknowns where $m=n+2$ is the number of grid points in each of the three dimensions including the two endpoints left out in the Gaussian quadrature rule. To express the BVP, we replace the leftmost and the rightmost equations by the Dirichlet BCs specified at the edges. This set of nonlinear algebraic equations can be solved efficiently by root-finding algorithms with one or more initial guesses, e.g., Newton's method and Brent's hybrid method\cite{NumericalRecipes,GSL}. The crudest initial guess can be a constant function. Better and faster solutions can be readily facilitated by using an earlier solution as the initial guess. This earlier solution can be one solved with fewer grid points and especially one that breaks a certain symmetry and thus certainly helps reach the distinct symmetry-related instanton paths. For the sake of further analysis of the instanton paths, any discrete solution obtained can be used to construct a continuous and differentiable function by cubic spline interpolation\cite{NumericalRecipes,GSL}.
\end{itemize}


\section{Details in symmetry analysis}\label{App:symmetry}
\subsection{Time-reversed solutions}\label{App:irreversibility}

As noted in Table~\ref{table1}, a pair of time-reversed paths of opposite gain and loss nature share the same total action. 
Let's show this by taking a look at the effective action Eq.~\eqref{eq:S_eff2} as an example. While other conventional terms are obviously invariant for the time-reversed path $\vec\sigma(-\tau)$, the temporally nonlocal $\mathcal{S}_\mathrm{i}$ in Eq.~\eqref{eq:S_i} makes no exception
\begin{equation}
\begin{split}
    &\mathcal{S}_\mathrm{i}[\vec\sigma(-\tau)] \\
    &=
    \frac{1}{T} \int_{-T}^{T} {\dd\tau\dd\tau' \sum_{i}{ \sigma_{i}(-\tau)\sigma_{i}(-\tau') K_i(\tau-\tau') }  } \\
    &= \frac{1}{T} \int_{T}^{-T} {(-\dd\tau)(-\dd\tau') \sum_{i}{ \sigma_{i}(\tau)\sigma_{i}(\tau') K_i(\tau'-\tau) }  } \\
    &= \frac{1}{T} \int_{-T}^{T} {\dd\tau\dd\tau' \sum_{i}{ \sigma_{i}(\tau)\sigma_{i}(\tau') K_i(\tau'-\tau) }  } \\
    &= \frac{1}{T} \int_{-T}^{T} {\dd\tau'\dd\tau \sum_{i}{ \sigma_{i}(\tau')\sigma_{i}(\tau) K_i(\tau-\tau') }  } \\
    &= \mathcal{S}_\mathrm{i}[\vec\sigma(\tau)].
\end{split}
\end{equation}
This is not surprising since the same conclusion holds as well with $\vec{\sigma}(-\tau),\vec{q}(-\tau)$ for the equivalent and original action in Eq.~\eqref{eq:gaussianInt} that corresponds to Eq.~\eqref{eq:H_all}.

The system's energy at an instant of imaginary time is $E(\tau)=\frac{1}{2}\sum_i{m_i {\dot Q_i}^2}-V(\vec{Q})$, which is certainly not conserved. Therefore, from Eq.~\eqref{eq:EOM} the rate of energy variation is
\begin{equation}\label{eq:dotE}
    \dot{E}(\tau) = \sum_i d_i \dot{Q}_i  \int_{- T}^{ T}{ \dd x' \frac{(Q_{i\tau}-Q_{i\tau'})}{(\tau-\tau')^2} }.
\end{equation}
Using these expressions, one can examine the gain and loss behavior of obtained instanton solutions.

\subsection{QD instanton}\label{App:QD}
We explain here why QD instantons are not able to exhibit Berry phase interference. The crucial observation is that any dissipative QD instanton outward from and back to a metastable $\vec{Q}_0$, composed of two segments of interchanged BCs, must be $\mathcal{T}$-symmetric. This is simply because a QD instanton comprising two 
$\mathcal{T}$-related segments in the first row of Table~\ref{table1} has the minimal action. On the other hand, a loop path (two segments from a column of Table~\ref{table1}) encircling any Berry flux is energetically unfavorable than the previous one. 
Since $\mathcal{T}\vec{Q}(\tau)$ exactly retraces back $\vec{Q}(\tau)$ and cancels any phase accumulated, dissipative QD instantons never bear any interference. This has been previously only noticed in the absence of dissipation\cite{Garg1993}.

\subsection{Mirror symmetry protection}\label{App:mirror}
The first line of Table~\ref{table2} can be verified in Eq.~\eqref{eq:EOM}. In general, the instanton path $\vec{Q}'(\tau)$ with energy gain, accompanied by the identical action as $\vec{Q}(\tau)$ has, consists of $Q_i(-\tau)$ in the $i$-direction when $Q_i(-T)=Q_i(T)$ regardless of the existence of mirror symmetry $\mathcal{M}_i$ and $-Q_j(-\tau)$ in the $j$-direction when $Q_j(-T)=-Q_j(T)$ and there is mirror symmetry $\mathcal{M}_j$. This complies with the general conditions for CT in Append.~\ref{App:landscape} and originates from the mirror symmetry $\mathcal{M}_j$. Note that one can translate and rotate the coordinate system to have $Q_i(-T)=Q_i(T)$ in two directions and $Q_j(-T)=-Q_j(T)$ in the other direction of two degenerate local minima. However, this alternative path $\vec{Q}'(\tau)$ does not contribute to any interference effect to $\vec{Q}(\tau)$. Let's exemplify with two case discussed in Append.~\ref{App:landscape}. For $\vec{w}=w_1\hat{1}+w_2\hat{2}$, since we have $\cos\theta'(\tau)=-\cos\theta(-\tau)$ and $\dot\phi'(\tau)=-\dot\phi(-\tau)$, the Berry phase term $\mathcal{S}_\mathrm{\Phi}$ in Eq.~\eqref{eq:spinBerry} differs only by an integral of the total derivative $\dot\phi$, which vanishes as dictated by the BC. The proof becomes even simpler for the equivalent but 
rotated case $\vec{w}=w_2\hat{2}+w_3\hat{3}$ adopted in Sec.~\ref{Sec:symmetry}, where $\cos\theta'(\tau)=\cos\theta(-\tau),\dot\phi'(\tau)=\dot\phi(-\tau)$. This absence of interference between $\vec{Q}(\tau)$ and $\vec{Q}'(\tau)$ is in a way an effect of the mirror symmetry.

Nonetheless, while the above discussion still holds, the situation alters when there is one mirror symmetry more. For an instanton solution $\vec{Q}(\tau)$, there can be another equally possible instanton path $\vec{Q}^\mathrm{r}(\tau)$ that differs in the $i$-direction by $-Q_i(\tau)$ as a solution to Eq.~\eqref{eq:EOM} when $Q_i(-T)=Q_i(T)=0,\alpha_j>\alpha_{i,k}$ and there are two mirror symmetries $\mathcal{M}_i,\mathcal{M}_j$ with mutually unequal $\alpha_{i,j,k}$. Note that one cannot arbitrarily translate the coordinate system to accommodate this special boundary position since the quadratic and monopolar potentials in Eq.~\eqref{eq:potential} in general have distinct inversion centers when $\vec{w}\neq0$.
For instance, when $\vec{w}=w\hat{3}$ as discussed in Append.~\ref{App:landscape}, one has mirror symmetries $\mathcal{M}_1,\mathcal{M}_2$. Any instanton path $\vec{Q}(\tau)$ also has a symmetry related pair $\vec{Q}^\mathrm{r}(\tau)=\mathcal{M}_2 \vec{Q}(\tau)=(Q_1(\tau),-Q_2(\tau),Q_3(\tau))$ as another path connecting the endpoints related by $\mathcal{M}_1$, which gives $\cos\theta^\mathrm{r}(\tau)=\cos\theta(\tau),\dot\phi^\mathrm{r}(\tau)=-\dot\phi(\tau)$ and hence a reversed Berry phase factor $\Phi^\mathrm{r}=-\Phi$ where $\Phi$ is for $\vec{Q}(\tau)$. In the absence of the extra mirror symmetry, there is only one single instanton path $\vec{Q}(\tau)$ and its pair $\vec{Q}'(\tau)$ with no phase difference. 
Now the CT tunnel amplitude is calculated as shown in Eq.~\eqref{eq:CT_main},
where we have used the fact that $\mathcal{S}_\mathrm{Q}[\vec{Q}(\tau)]=\mathcal{S}_\mathrm{Q}[\vec{Q}^\mathrm{r}(\tau)]\equiv\mathcal{S}_\mathrm{Q}\,,\mathcal{S}_\mathrm{\Phi}[\vec{Q}(\tau)]=-\mathcal{S}_\mathrm{\Phi}[\vec{Q}^\mathrm{r}(\tau)]\equiv\mathcal{S}_\mathrm{\Phi}=-\mathcal{S}_\mathrm{\Phi}^\mathrm{r}=\ii \Phi$. 
The CT tunnel splitting $\Delta\propto A_0$, which can be seen from Eq.~\eqref{eq:transitionAmp2}.


\section{Analytical analysis of instanton solutions}\label{App:PhaseDiagram}
\begin{figure*}[hbt]
\includegraphics[width=0.8\textwidth]{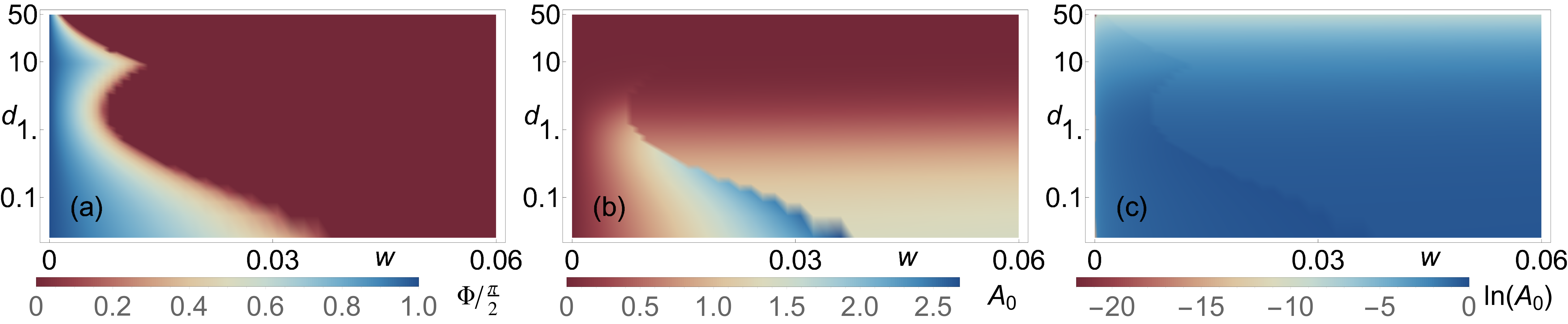}
\caption{Phase diagrams of (a) instanton Berry phase $\Phi$ and (b) coherent tunnel amplitude $A_0$ and (c) logarithmic scaled $\ln{A_0}$ against potential bias $w$ and logarithmic scaled dissipation strength $d$. Instantons are solved with settings same as Fig.~\ref{Fig:PhaseDiagram}, scanning a $50\times 50$-mesh of the $wd$-plane that is denser for smaller $d$.}
\label{Fig:FullPhaseDiagram}
\end{figure*} 
We provide some analytical analysis to understand the behavior of the instantons in the phase diagram Fig.~\ref{Fig:PhaseDiagram} where we have $\vec{w}=w\hat{3}$.
\begin{itemize}
    \item $w=0$
\begin{itemize}
    \item $\alpha_1>\alpha_2>\alpha_3$
    This directly corresponds to the calculation shown in Sec.~\ref{Sec:phasediagram}.
    The instanton path must lie in the $Q_1Q_2$-plane, because one can rotate, with respect to the $Q_1$-axis, any path with finite $Q_3$ component into the $Q_1Q_2$-plane as per the homogeneous 
    BCs in $Q_{2,3}$ and achieve a smaller action in Eq.~\eqref{eq:S_eff3}, which holds regardless of the strength of the dissipation. 
    Therefore, the instanton paths always come with full topological suppression due to $\Phi=\pi/2$ as long as the system spontaneously exhibits finite $Q_2$, which is what one observes up to very large $d$ as shown in Fig.~\ref{Fig:FullPhaseDiagram}.
    \item $\alpha_1>\alpha_3>\alpha_2$
    Same as the previous one up to an exchange of coordinate $2,3$ as one has both $\mathcal{M}_2$ and $\mathcal{M}_3$ mirror symmetries. 
\end{itemize}
    \item $w>0$
\begin{itemize}
    \item $\alpha_1>\alpha_2>\alpha_3$
    This directly corresponds to the calculation shown in Sec.~\ref{Sec:phasediagram}. The previous rotation argument no longer applies since finite $w$ renders a rotated path not fulfilling the instanton's inhomogeneous 
    BC in $Q_3$. Therefore, the instanton path will acquire a finite $Q_3$ component. While the instanton alters to be restricted in the $Q_1Q_3$-plane outside the critical curve $f$, it is in general \textit{not} within any certain 2D plane inside $f$
    , which is consistent with the nonconservation of angular momentum due to the noncentral force from $V$ in Eq.~\eqref{eq:EOM}. This is also verified in the numerical solutions using the least square matrix eigenvalues of points along an instanton path, although barely discernible in Fig.~\ref{Fig:Paths}. Also note that $\alpha_2=\alpha_3$ bears a degenerate minima ring as aforementioned in Append.~\ref{App:landscape}, invalidating the double-minima discussions so far. Indeed, as $\alpha_2\rightarrow\alpha_3+0^+$, any $Q_2\neq0$ path eventually disappears and the critical curve $f$ shrinks inwards, which connects to the following case.
    \item $\alpha_1>\alpha_3>\alpha_2$
    Paths with finite $Q_2$ are not energetically favorable and because of the finite $w_3$, there is only one single instanton path in the upper $Q_1Q_3$-plane and thus $\Phi=0$.
\end{itemize}
\end{itemize}




For the region of larger $d$ not comparable to other system parameters, the phase boundary curve $f$ possesses a complex shape, because within the finite phase region inside $f$, $\Phi(w>0,d)$ does not vary monotonically with $d$, as shown in Fig.~\ref{Fig:FullPhaseDiagram}(a). At finite but not too large $w$, it first decreases until a stationary point, then increases until a second stationary point, and finally decreases to zero. Instead of having the first stationary point, for slightly larger $w$ it may show an intermediate region of $\Phi=0$ at not too large $d$, which means an inflection point and a convex hump with cusp outward on the $f$ curve. Usually, for $d$ beyond this inflection point higher than $d=1$, the instanton solution begins to deform and, dependent on a certain direction's BC is parity even or odd, asymptotically approach in that direction a $\mathcal{T}$-symmetric or  $\mathcal{T}$-antisymmetric form without net energy change, as the nonlocal interaction term eventually outweighs the potential $V$ in Eq.~\eqref{eq:EOM}. On the other hand, for larger and larger $w$ beyond $f$ and towards the edge of the CT condition $w=(\alpha_1-\alpha_3)/2$ mentioned in Sec.~\ref{Sec:EOM}, the BC dictates that the range roved by an instanton reduces towards zero and stays local and high in the $Q_1Q_3$-plane relative to the monopole at the origin. Within the instanton path, the potential $V$ thereby barely varies around the minimum value. Again, we are led to have the potential $V$ suppressed and asymptotically reach $\mathcal{T}$-(anti)symmetric behavior. 

To understand this inclination towards (anti)symmetrization, we can thus consider Eq.~\eqref{eq:EOM} without the potential term, which has a linear integro-differential operator of even parity. This means that from any solution $Q_i(\tau)$ in the $i$-direction one can construct parity even/odd solutions $Q_i^{\mathrm{e/o}}(\tau)=(Q_i(\tau)\pm Q_i(-\tau))/2$, which can all be made consistent with the even/odd BCs in $Q_3$/$Q_1$ while $Q_2$ satisfies both. From the linearity and the physical nature, the uniqueness of solution guarantees $Q_3(\tau)$/$Q_1(\tau)$ is even/odd and $Q_2$ is parity definite or zero. \mycomment{Note that the kinetic term is nonnegligible since the integral equation itself does not have valid solutions\red{??}}This holds even when the kinetic term is also negligible. However, it is never legitimate to completely drop the potential term, otherwise the three components $Q_{1,2,3}$ are totally decoupled and one is left with trivial solutions, i.e., the potential $V$ is in a way nonperturbative to the system. The competition due to larger $d$ between the nonlinear $V$ and the trend of (anti)symmetrization generates the complex phase boundary with the cusped hump structure, although eventually at large enough $d$ the phase $\Phi$ will tend to zero at any finite $w$ as shown.

Across the phase boundary $f$ curve in Fig.~\ref{Fig:FullPhaseDiagram}(a), the corresponding $A_0$-jump always exists, although it eventually gets smoothed out at large enough $d$ since large enough $\mathcal{S}_\mathrm{Q}$ exponentially suppresses the jump, as shown in Fig.~\ref{Fig:FullPhaseDiagram}(b). To reassure, we show a logarithmic plot in Fig.~\ref{Fig:FullPhaseDiagram}(c), where the jump becomes more visible. In summary, beyond the phase diagram Fig.~\ref{Fig:PhaseDiagram}, for larger $d$ that is less and less comparable to $m,w,\vec\alpha$ typically equal or less than unity, 
$\Phi$ can vary nonmonotonically 
although eventually vanishes
\mycomment{\cite{SM}}. The complexities originate from that when potential $V$ becomes less important the remaining parity-even linear integro-differential operator causes the instanton to (anti)symmetrize in $\tau$, although $V$ is still crucial to nonlinearly couple three directions. 
We describe here this large-$d$ behavior for completeness, probably of less physical interest as the modeling could become less reliable towards realistic situations.


\begin{widetext}
\section{Dissipative instanton gas tunnelling}\label{App:gas}
One might generalize the instantons discussed so far to a 1D dissipative instanton gas with multiple instantons and interaction as well. For instance, although dissipation has been included in the action of single instantons, it still exists between instanton events at different times, which gives a 1D logarithmic interaction $C_{ij}(|\tau_i-\tau_j|)=-c_{ij}\ln{|\tau_i-\tau_j|}$ in the dilute limit\cite{Schmid1983}. Incorporating this between (anti)instantons leads to the grand-canonical partition function, the total transition amplitude from $\vec{Q}_-$ to $\vec{Q}_+$ or back to $\vec{Q}_-$ is given by
\begin{equation}\label{eq:gasZ}
\begin{split}
    A(\vec{Q}_-,\vec{Q}_\mp) = \mathcal{Z}_\mathrm{IG} 
    &= \sum_{n=0}^\infty \sum_{\{p_i=\pm1,\xi_i=\pm1\}} \int_{-T}^ T\dd\tau_1
\int_{-T}^{\tau_1}\dd\tau_2
\cdots
\int_{-T}^{\tau_{n+m-1}}\dd\tau_{n+m} \ee^{-\left[ \sum_{i<j}^{n+m} q_iq_j C_{ij}(|\tau_i-\tau_j|) + \sum_i (\mathcal{S}_\mathrm{Q}+\ii q_i\xi_i\Phi) \right]} \\
&= \sum_{n=0}^\infty \int_{-T}^ T\dd\tau_1
\int_{-T}^{\tau_1}\dd\tau_2
\cdots
\int_{-T}^{\tau_{n+m-1}}\dd\tau_{n+m} y^{n+m} \ee^{-\left[ \sum_{i<j}^{n+m} q_iq_j C(|\tau_i-\tau_j|)  \right]},
\end{split}
\end{equation}
where $m=n+\frac{1\mp1}{2}$, the charges $\xi_i=\pm1,p_i=\pm1,q_i=(-1)^{i+1}$ denote for the $i$th instanton respectively whether it is $\vec{Q}^\mathrm{r}$ instanton, whether it is $\vec{Q}'$ instanton although this does not affect the single-instanton action directly, and the direction of motion (instanton/antiinstanton). In principle, instanton events centered at $\tau_i$ and $\tau_j$ could have their interaction coefficient $c_{ij}$ dependent on the respective instanton type (among the possible four) as the overlap integral differs, i.e., dependent on the charges $\xi_i,p_i$, hence the subscript of $c_{ij}$. In this case, the summation in the first line of Eq.~\eqref{eq:gasZ} renders the fugacity inseparable from the interaction, which is a more complex instanton gas than usual cases like the Kondo analogy in Sec.~\ref{Sec:gas}.

In the dilute limit, assuming that the instanton interaction does not depend on the charges $\xi_i,p_i$, i.e., $c_{ij}=c$, one can perform the summation on the gas fugacity due to the instanton core energy as shown in the second line of Eq.~\eqref{eq:gasZ}, leading to $y=2K\ee^{-\mathcal{S}_\mathrm{core}}$ with $\mathcal{S}_\mathrm{core}=\mathcal{S}_\mathrm{Q}-\ln{(2\cos{\Phi})}$ (see the noninteracting case below). 
An interacting neutral Coulomb gas can be mapped to the sine-Gordon model\mycomment{\cite{Samuel1978}, which has been used to study the transition from diffusion to localization in a frictional sine-Gordon model\cite{Schmid1983}}. 
However, the reverse procedure is not applicable here since the time ordering of the instantons is in general unremovable. In other words, the failure partially lies in the invalidity of the following transformation
$\int_{-T}^ T\dd\tau_1
\int_{-T}^{\tau_1}\dd\tau_2
\cdots
\int_{-T}^{\tau_{n+m-1}}\dd\tau_{n+m} f(\{|\tau_i-\tau_j|\})
\neq
\frac{1}{n!m!}
\int_{-T}^ T\dd\tau_1
\int_{-T}^{ T}\dd\tau_2
\cdots
\int_{-T}^{ T}\dd\tau_{n+m} f(\{|\tau_i-\tau_j|\})$
where $f(\{|\tau_i-\tau_j|\})$ denotes a generic function dependent on all possible $|\tau_i-\tau_j|\,,i\ne j$.
Instead, it formally resembles the anisotropic Kondo problem. In this regard, $c=2\mathcal{K}$ where $\mathcal{K}=d\,[2Q_1(T)]^2/2$ determines whether the system flows to localization ($\mathcal{K}>1$) or not ($\mathcal{K}<1$) and $\mathcal{K}=1/2$ is the Toulouse limit\cite{Leggett1987}. To see this, in the anisotropic Kondo model 
\begin{equation}
    \mathcal{H}_\mathrm{AK}=v_F\sum_{k,\sigma}{kc_{k\sigma}^\dag c_{k\sigma}}+J_\parallel/4\,s_z\sum_\sigma{\sigma c_{\sigma}^\dag c_{\sigma}} + J_\perp/2\,(s_+c_{\downarrow}^\dag c_{\uparrow}+s_-c_{\uparrow}^\dag c_{\downarrow})
\end{equation}
we would have $y=\rho J_\perp/2,\mathcal{K}=(1-\rho J_\parallel),\rho=1/2\pi v_F$, where $\vec{s}$ is the lozalized spin. And the renormalization group equations are to the lowest order 
\begin{equation}\label{eq:RG}
\begin{split}
    \dd(1-\mathcal{K})/\dd\ln{\tau_c} = \mathcal{K}y^2/4 \\
    \dd y/\dd\ln{\tau_c} = (1-\mathcal{K})y
\end{split}
\end{equation}
where $1/\tau_c$ denotes the running high-energy cutoff.

We now consider the case of a noninteracting dissipative dilute instanton gas in more detail. By taking into account all possible numbers and configurations of instantons, 
\begin{equation}
    A(\vec{Q}_-,\vec{Q}_\mp) = \sum_{n=\mathrm{even}/\mathrm{odd}}{ K^n \int_{-T}^T \dd\tau_1 \int_{-T}^{\tau_1} \dd\tau_2 \cdots \int_{-T}^{\tau_{n-1}} \dd\tau_n A_n(\tau_1,\cdots,\tau_n) }
\end{equation}
where the fluctuation determinant $K$ is a prefactor of the exponential amplitude $A_{n}$ and can be calculated by taking into account the possible Goldstone mode in $\tau$-space\cite{\mycomment{Garg1992,}Weiss2012,Caldeira2014}. We do not evaluate $K$ here as it does not affect the exponential accuracy that we are mainly interested in.
The transition amplitude $A_{n} = A_{n}^\mathrm{inst} A_{n}^\mathrm{flct}$ with $n$ (anti)instantons consists of two parts.
The instanton part
\begin{equation}
    A_{n}^\mathrm{inst}(\tau_1,\cdots,\tau_n) = \sum_{\{p_i=\pm1,\xi_i=\pm 1\}} { \prod_i { \ee^{-(\mathcal{S}_\mathrm{Q}+\ii\,\xi_iq_i\Phi)} } }
    = \ee^{-n\mathcal{S}_\mathrm{Q}} \prod_i\sum_{p_i=\pm1,\xi_i=\pm 1}\ee^{-\ii\,\xi_iq_i\Phi} = (4 \ee^{-\mathcal{S}_\mathrm{Q}}\cos\Phi)^n = A_0^n.
\end{equation}
The quantum fluctuation part accounts for the harmonic fluctuation accumulated while sitting at the minima
    $A_{n}^\mathrm{flct} = \prod_i \ee^{-\omega(\tau_{i+1}-\tau_i)/2} = \ee^{-\omega T}$
wherein $\omega$ is frequency determined by the harmonic potential approximation $m\omega^2=V''(\vec{Q}_\pm)$\cite{CMFT}.
Therefore, we have
\begin{equation}\label{eq:transitionAmp2}
\begin{split}
    A(\vec{Q}_-,\vec{Q}_\mp) &= B\sum_{n=\mathrm{even}/\mathrm{odd}} K^n \ee^{-\omega T} A_0^n \int_{-T}^ T \dd\tau_1 \int_{-T}^{\tau_1} \dd\tau_2 \cdots \int_{-T}^{\tau_{n-1}} \dd\tau_n \\
    &= B\:\ee^{-\omega T} \sum_{n=\mathrm{even}/\mathrm{odd}} \frac{1}{n!} (2 K TA_0)^n 
    = B\ee^{-\omega T} 
    \begin{cases}
    \cosh{(2 K TA_0)} & n=\mathrm{even} \\
    \sinh{(2 K TA_0)} & n=\mathrm{odd}
    \end{cases}.
\end{split}
\end{equation}
Prefactor $B$ is introduced to account for the state overlap with the spontaneously formed even/odd state $\ket{e/o}$, i.e., $\braket{\vec{Q}_+|e}=\braket{\vec{Q}_-|e}$, $\braket{\vec{Q}_+|o}=-\braket{\vec{Q}_-|o}$, and $B/2=|\braket{\vec{Q}_\mp|e/o}|^2$.
It can then be compared to the same quantity calculated from the $\ket{e/o}$ state splitting 
\begin{equation}\label{eq:transitionAmp3}
    A(\vec{Q}_-,\vec{Q}_\mp) 
    = \bra{\vec{Q}_-} (\ee^{-(\omega-\Delta)T}\ket{e}\bra{e} + \ee^{-(\omega+\Delta)T}\ket{o}\bra{o}) \ket{\vec{Q}_\mp},
\end{equation}
from which we obtain the tunnel splitting $\Delta=KA_0$.
\end{widetext}


    
    
    



\bibliography{reference.bib}  

\begin{thebibliography}{73}%
\makeatletter
\providecommand \@ifxundefined [1]{%
 \@ifx{#1\undefined}
}%
\providecommand \@ifnum [1]{%
 \ifnum #1\expandafter \@firstoftwo
 \else \expandafter \@secondoftwo
 \fi
}%
\providecommand \@ifx [1]{%
 \ifx #1\expandafter \@firstoftwo
 \else \expandafter \@secondoftwo
 \fi
}%
\providecommand \natexlab [1]{#1}%
\providecommand \enquote  [1]{``#1''}%
\providecommand \bibnamefont  [1]{#1}%
\providecommand \bibfnamefont [1]{#1}%
\providecommand \citenamefont [1]{#1}%
\providecommand \href@noop [0]{\@secondoftwo}%
\providecommand \href [0]{\begingroup \@sanitize@url \@href}%
\providecommand \@href[1]{\@@startlink{#1}\@@href}%
\providecommand \@@href[1]{\endgroup#1\@@endlink}%
\providecommand \@sanitize@url [0]{\catcode `\\12\catcode `\$12\catcode
  `\&12\catcode `\#12\catcode `\^12\catcode `\_12\catcode `\%12\relax}%
\providecommand \@@startlink[1]{}%
\providecommand \@@endlink[0]{}%
\providecommand \url  [0]{\begingroup\@sanitize@url \@url }%
\providecommand \@url [1]{\endgroup\@href {#1}{\urlprefix }}%
\providecommand \urlprefix  [0]{URL }%
\providecommand \Eprint [0]{\href }%
\providecommand \doibase [0]{https://doi.org/}%
\providecommand \selectlanguage [0]{\@gobble}%
\providecommand \bibinfo  [0]{\@secondoftwo}%
\providecommand \bibfield  [0]{\@secondoftwo}%
\providecommand \translation [1]{[#1]}%
\providecommand \BibitemOpen [0]{}%
\providecommand \bibitemStop [0]{}%
\providecommand \bibitemNoStop [0]{.\EOS\space}%
\providecommand \EOS [0]{\spacefactor3000\relax}%
\providecommand \BibitemShut  [1]{\csname bibitem#1\endcsname}%
\let\auto@bib@innerbib\@empty
\bibitem [{\citenamefont {Berry}(1984)}]{Berry1984}%
  \BibitemOpen
  \bibfield  {author} {\bibinfo {author} {\bibfnamefont {M.~V.}\ \bibnamefont
  {Berry}},\ }\href {https://doi.org/10.1098/rspa.1984.0023} {\bibfield
  {journal} {\bibinfo  {journal} {Proceedings of the Royal Society A:
  Mathematical, Physical and Engineering Sciences}\ }\textbf {\bibinfo {volume}
  {392}},\ \bibinfo {pages} {45} (\bibinfo {year} {1984})}\BibitemShut
  {NoStop}%
\bibitem [{\citenamefont {Bohm}\ \emph {et~al.}(2003)\citenamefont {Bohm},
  \citenamefont {Mostafazadeh}, \citenamefont {Koizumi}, \citenamefont {Niu},\
  and\ \citenamefont {Zwanziger}}]{Bohm2003}%
  \BibitemOpen
  \bibfield  {author} {\bibinfo {author} {\bibfnamefont {A.}~\bibnamefont
  {Bohm}}, \bibinfo {author} {\bibfnamefont {A.}~\bibnamefont {Mostafazadeh}},
  \bibinfo {author} {\bibfnamefont {H.}~\bibnamefont {Koizumi}}, \bibinfo
  {author} {\bibfnamefont {Q.}~\bibnamefont {Niu}},\ and\ \bibinfo {author}
  {\bibfnamefont {J.}~\bibnamefont {Zwanziger}},\ }\href
  {https://doi.org/10.1007/978-3-662-10333-3} {\emph {\bibinfo {title} {The
  Geometric Phase in Quantum Systems}}}\ (\bibinfo  {publisher} {Springer
  Berlin Heidelberg},\ \bibinfo {address} {Heidelberg},\ \bibinfo {year}
  {2003})\BibitemShut {NoStop}%
\bibitem [{\citenamefont {Xiao}\ \emph {et~al.}(2010)\citenamefont {Xiao},
  \citenamefont {Chang},\ and\ \citenamefont {Niu}}]{Xiao2010}%
  \BibitemOpen
  \bibfield  {author} {\bibinfo {author} {\bibfnamefont {D.}~\bibnamefont
  {Xiao}}, \bibinfo {author} {\bibfnamefont {M.-C.}\ \bibnamefont {Chang}},\
  and\ \bibinfo {author} {\bibfnamefont {Q.}~\bibnamefont {Niu}},\ }\href
  {https://doi.org/10.1103/revmodphys.82.1959} {\bibfield  {journal} {\bibinfo
  {journal} {Reviews of Modern Physics}\ }\textbf {\bibinfo {volume} {82}},\
  \bibinfo {pages} {1959} (\bibinfo {year} {2010})}\BibitemShut {NoStop}%
\bibitem [{\citenamefont {Nagaosa}\ \emph {et~al.}(2010)\citenamefont
  {Nagaosa}, \citenamefont {Sinova}, \citenamefont {Onoda}, \citenamefont
  {MacDonald},\ and\ \citenamefont {Ong}}]{Nagaosa2010}%
  \BibitemOpen
  \bibfield  {author} {\bibinfo {author} {\bibfnamefont {N.}~\bibnamefont
  {Nagaosa}}, \bibinfo {author} {\bibfnamefont {J.}~\bibnamefont {Sinova}},
  \bibinfo {author} {\bibfnamefont {S.}~\bibnamefont {Onoda}}, \bibinfo
  {author} {\bibfnamefont {A.~H.}\ \bibnamefont {MacDonald}},\ and\ \bibinfo
  {author} {\bibfnamefont {N.~P.}\ \bibnamefont {Ong}},\ }\href
  {https://doi.org/10.1103/revmodphys.82.1539} {\bibfield  {journal} {\bibinfo
  {journal} {Rev. Mod. Phys.}\ }\textbf {\bibinfo {volume} {82}},\ \bibinfo
  {pages} {1539} (\bibinfo {year} {2010})}\BibitemShut {NoStop}%
\bibitem [{\citenamefont {Sinova}\ \emph {et~al.}(2015)\citenamefont {Sinova},
  \citenamefont {Valenzuela}, \citenamefont {Wunderlich}, \citenamefont
  {Back},\ and\ \citenamefont {Jungwirth}}]{Sinova2015}%
  \BibitemOpen
  \bibfield  {author} {\bibinfo {author} {\bibfnamefont {J.}~\bibnamefont
  {Sinova}}, \bibinfo {author} {\bibfnamefont {S.~O.}\ \bibnamefont
  {Valenzuela}}, \bibinfo {author} {\bibfnamefont {J.}~\bibnamefont
  {Wunderlich}}, \bibinfo {author} {\bibfnamefont {C.~H.}\ \bibnamefont
  {Back}},\ and\ \bibinfo {author} {\bibfnamefont {T.}~\bibnamefont
  {Jungwirth}},\ }\href {https://doi.org/10.1103/revmodphys.87.1213} {\bibfield
   {journal} {\bibinfo  {journal} {Reviews of Modern Physics}\ }\textbf
  {\bibinfo {volume} {87}},\ \bibinfo {pages} {1213} (\bibinfo {year}
  {2015})}\BibitemShut {NoStop}%
\bibitem [{\citenamefont {Xiao}\ \emph {et~al.}(2012)\citenamefont {Xiao},
  \citenamefont {Liu}, \citenamefont {Feng}, \citenamefont {Xu},\ and\
  \citenamefont {Yao}}]{Xiao2012}%
  \BibitemOpen
  \bibfield  {author} {\bibinfo {author} {\bibfnamefont {D.}~\bibnamefont
  {Xiao}}, \bibinfo {author} {\bibfnamefont {G.-B.}\ \bibnamefont {Liu}},
  \bibinfo {author} {\bibfnamefont {W.}~\bibnamefont {Feng}}, \bibinfo {author}
  {\bibfnamefont {X.}~\bibnamefont {Xu}},\ and\ \bibinfo {author}
  {\bibfnamefont {W.}~\bibnamefont {Yao}},\ }\href
  {https://doi.org/10.1103/physrevlett.108.196802} {\bibfield  {journal}
  {\bibinfo  {journal} {Physical Review Letters}\ }\textbf {\bibinfo {volume}
  {108}},\ \bibinfo {pages} {196802} (\bibinfo {year} {2012})}\BibitemShut
  {NoStop}%
\bibitem [{\citenamefont {Cao}\ \emph {et~al.}(2012)\citenamefont {Cao},
  \citenamefont {Wang}, \citenamefont {Han}, \citenamefont {Ye}, \citenamefont
  {Zhu}, \citenamefont {Shi}, \citenamefont {Niu}, \citenamefont {Tan},
  \citenamefont {Wang}, \citenamefont {Liu},\ and\ \citenamefont
  {Feng}}]{Cao2012}%
  \BibitemOpen
  \bibfield  {author} {\bibinfo {author} {\bibfnamefont {T.}~\bibnamefont
  {Cao}}, \bibinfo {author} {\bibfnamefont {G.}~\bibnamefont {Wang}}, \bibinfo
  {author} {\bibfnamefont {W.}~\bibnamefont {Han}}, \bibinfo {author}
  {\bibfnamefont {H.}~\bibnamefont {Ye}}, \bibinfo {author} {\bibfnamefont
  {C.}~\bibnamefont {Zhu}}, \bibinfo {author} {\bibfnamefont {J.}~\bibnamefont
  {Shi}}, \bibinfo {author} {\bibfnamefont {Q.}~\bibnamefont {Niu}}, \bibinfo
  {author} {\bibfnamefont {P.}~\bibnamefont {Tan}}, \bibinfo {author}
  {\bibfnamefont {E.}~\bibnamefont {Wang}}, \bibinfo {author} {\bibfnamefont
  {B.}~\bibnamefont {Liu}},\ and\ \bibinfo {author} {\bibfnamefont
  {J.}~\bibnamefont {Feng}},\ }\href {https://doi.org/10.1038/ncomms1882}
  {\bibfield  {journal} {\bibinfo  {journal} {Nature Communications}\ }\textbf
  {\bibinfo {volume} {3}},\ \bibinfo {pages} {887} (\bibinfo {year}
  {2012})}\BibitemShut {NoStop}%
\bibitem [{\citenamefont {Xu}\ \emph {et~al.}(2014)\citenamefont {Xu},
  \citenamefont {Yao}, \citenamefont {Xiao},\ and\ \citenamefont
  {Heinz}}]{Xu2014}%
  \BibitemOpen
  \bibfield  {author} {\bibinfo {author} {\bibfnamefont {X.}~\bibnamefont
  {Xu}}, \bibinfo {author} {\bibfnamefont {W.}~\bibnamefont {Yao}}, \bibinfo
  {author} {\bibfnamefont {D.}~\bibnamefont {Xiao}},\ and\ \bibinfo {author}
  {\bibfnamefont {T.~F.}\ \bibnamefont {Heinz}},\ }\href
  {https://doi.org/10.1038/nphys2942} {\bibfield  {journal} {\bibinfo
  {journal} {Nature Physics}\ }\textbf {\bibinfo {volume} {10}},\ \bibinfo
  {pages} {343} (\bibinfo {year} {2014})}\BibitemShut {NoStop}%
\bibitem [{\citenamefont {Ren}\ \emph {et~al.}(2016)\citenamefont {Ren},
  \citenamefont {Qiao},\ and\ \citenamefont {Niu}}]{DiracFermion2}%
  \BibitemOpen
  \bibfield  {author} {\bibinfo {author} {\bibfnamefont {Y.}~\bibnamefont
  {Ren}}, \bibinfo {author} {\bibfnamefont {Z.}~\bibnamefont {Qiao}},\ and\
  \bibinfo {author} {\bibfnamefont {Q.}~\bibnamefont {Niu}},\ }\href
  {https://doi.org/10.1088/0034-4885/79/6/066501} {\bibfield  {journal}
  {\bibinfo  {journal} {Rep. Prog. Phys.}\ }\textbf {\bibinfo {volume} {79}},\
  \bibinfo {pages} {066501} (\bibinfo {year} {2016})}\BibitemShut {NoStop}%
\bibitem [{\citenamefont {Hasan}\ and\ \citenamefont {Kane}(2010)}]{Topo1}%
  \BibitemOpen
  \bibfield  {author} {\bibinfo {author} {\bibfnamefont {M.~Z.}\ \bibnamefont
  {Hasan}}\ and\ \bibinfo {author} {\bibfnamefont {C.~L.}\ \bibnamefont
  {Kane}},\ }\href {https://doi.org/10.1103/RevModPhys.82.3045} {\bibfield
  {journal} {\bibinfo  {journal} {Rev. Mod. Phys.}\ }\textbf {\bibinfo {volume}
  {82}},\ \bibinfo {pages} {3045} (\bibinfo {year} {2010})}\BibitemShut
  {NoStop}%
\bibitem [{\citenamefont {Qi}\ and\ \citenamefont {Zhang}(2011)}]{Topo2}%
  \BibitemOpen
  \bibfield  {author} {\bibinfo {author} {\bibfnamefont {X.-L.}\ \bibnamefont
  {Qi}}\ and\ \bibinfo {author} {\bibfnamefont {S.-C.}\ \bibnamefont {Zhang}},\
  }\href {https://doi.org/10.1103/RevModPhys.83.1057} {\bibfield  {journal}
  {\bibinfo  {journal} {Rev. Mod. Phys.}\ }\textbf {\bibinfo {volume} {83}},\
  \bibinfo {pages} {1057} (\bibinfo {year} {2011})}\BibitemShut {NoStop}%
\bibitem [{\citenamefont {Yan}\ and\ \citenamefont
  {Felser}(2017)}]{ReviewYan2017}%
  \BibitemOpen
  \bibfield  {author} {\bibinfo {author} {\bibfnamefont {B.}~\bibnamefont
  {Yan}}\ and\ \bibinfo {author} {\bibfnamefont {C.}~\bibnamefont {Felser}},\
  }\href {https://doi.org/10.1146/annurev-conmatphys-031016-025458} {\bibfield
  {journal} {\bibinfo  {journal} {Annual Review of Condensed Matter Physics}\
  }\textbf {\bibinfo {volume} {8}},\ \bibinfo {pages} {337} (\bibinfo {year}
  {2017})}\BibitemShut {NoStop}%
\bibitem [{\citenamefont {{Armitage}}\ \emph {et~al.}(2018)\citenamefont
  {{Armitage}}, \citenamefont {{Mele}},\ and\ \citenamefont
  {{Vishwanath}}}]{WeylDiracReview}%
  \BibitemOpen
  \bibfield  {author} {\bibinfo {author} {\bibfnamefont {N.~P.}\ \bibnamefont
  {{Armitage}}}, \bibinfo {author} {\bibfnamefont {E.~J.}\ \bibnamefont
  {{Mele}}},\ and\ \bibinfo {author} {\bibfnamefont {A.}~\bibnamefont
  {{Vishwanath}}},\ }\href {https://doi.org/10.1103/RevModPhys.90.015001}
  {\bibfield  {journal} {\bibinfo  {journal} {Rev. Mod. Phys.}\ }\textbf
  {\bibinfo {volume} {90}},\ \bibinfo {pages} {015001} (\bibinfo {year}
  {2018})}\BibitemShut {NoStop}%
\bibitem [{\citenamefont {Basov}\ \emph {et~al.}(2017)\citenamefont {Basov},
  \citenamefont {Averitt},\ and\ \citenamefont {Hsieh}}]{Basov2017}%
  \BibitemOpen
  \bibfield  {author} {\bibinfo {author} {\bibfnamefont {D.~N.}\ \bibnamefont
  {Basov}}, \bibinfo {author} {\bibfnamefont {R.~D.}\ \bibnamefont {Averitt}},\
  and\ \bibinfo {author} {\bibfnamefont {D.}~\bibnamefont {Hsieh}},\ }\href
  {https://doi.org/10.1038/nmat5017} {\bibfield  {journal} {\bibinfo  {journal}
  {Nat. Mater.}\ }\textbf {\bibinfo {volume} {16}},\ \bibinfo {pages} {1077}
  (\bibinfo {year} {2017})}\BibitemShut {NoStop}%
\bibitem [{\citenamefont {Tokura}\ \emph {et~al.}(2017)\citenamefont {Tokura},
  \citenamefont {Kawasaki},\ and\ \citenamefont {Nagaosa}}]{Tokura2017}%
  \BibitemOpen
  \bibfield  {author} {\bibinfo {author} {\bibfnamefont {Y.}~\bibnamefont
  {Tokura}}, \bibinfo {author} {\bibfnamefont {M.}~\bibnamefont {Kawasaki}},\
  and\ \bibinfo {author} {\bibfnamefont {N.}~\bibnamefont {Nagaosa}},\ }\href
  {https://doi.org/10.1038/nphys4274} {\bibfield  {journal} {\bibinfo
  {journal} {Nat. Phys.}\ }\textbf {\bibinfo {volume} {13}},\ \bibinfo {pages}
  {1056} (\bibinfo {year} {2017})}\BibitemShut {NoStop}%
\bibitem [{\citenamefont {Keimer}\ and\ \citenamefont
  {Moore}(2017)}]{Keimer2017}%
  \BibitemOpen
  \bibfield  {author} {\bibinfo {author} {\bibfnamefont {B.}~\bibnamefont
  {Keimer}}\ and\ \bibinfo {author} {\bibfnamefont {J.~E.}\ \bibnamefont
  {Moore}},\ }\href {https://doi.org/10.1038/nphys4302} {\bibfield  {journal}
  {\bibinfo  {journal} {Nature Physics}\ }\textbf {\bibinfo {volume} {13}},\
  \bibinfo {pages} {1045} (\bibinfo {year} {2017})}\BibitemShut {NoStop}%
\bibitem [{\citenamefont {Zurek}(2003)}]{Zurek2003}%
  \BibitemOpen
  \bibfield  {author} {\bibinfo {author} {\bibfnamefont {W.~H.}\ \bibnamefont
  {Zurek}},\ }\href {https://doi.org/10.1103/revmodphys.75.715} {\bibfield
  {journal} {\bibinfo  {journal} {Reviews of Modern Physics}\ }\textbf
  {\bibinfo {volume} {75}},\ \bibinfo {pages} {715} (\bibinfo {year}
  {2003})}\BibitemShut {NoStop}%
\bibitem [{\citenamefont {Dattagupta}\ and\ \citenamefont
  {Puri}(2004)}]{Dattagupta2004}%
  \BibitemOpen
  \bibfield  {author} {\bibinfo {author} {\bibfnamefont {S.}~\bibnamefont
  {Dattagupta}}\ and\ \bibinfo {author} {\bibfnamefont {S.}~\bibnamefont
  {Puri}},\ }\href {https://doi.org/10.1007/978-3-662-06758-1} {\emph {\bibinfo
  {title} {Dissipative Phenomena in Condensed Matter}}}\ (\bibinfo  {publisher}
  {Springer Berlin Heidelberg},\ \bibinfo {address} {Heidelberg},\ \bibinfo
  {year} {2004})\BibitemShut {NoStop}%
\bibitem [{\citenamefont {Weiss}(2012)}]{Weiss2012}%
  \BibitemOpen
  \bibfield  {author} {\bibinfo {author} {\bibfnamefont {U.}~\bibnamefont
  {Weiss}},\ }\href {https://doi.org/10.1142/8334} {\emph {\bibinfo {title}
  {Quantum Dissipative Systems}}}\ (\bibinfo  {publisher} {World Scientific
  Publishing Company},\ \bibinfo {address} {Singapore},\ \bibinfo {year}
  {2012})\BibitemShut {NoStop}%
\bibitem [{\citenamefont {Caldeira}(2014)}]{Caldeira2014}%
  \BibitemOpen
  \bibfield  {author} {\bibinfo {author} {\bibfnamefont {A.~O.}\ \bibnamefont
  {Caldeira}},\ }\href {https://doi.org/10.1017/cbo9781139035439} {\emph
  {\bibinfo {title} {An Introduction to Macroscopic Quantum Phenomena and
  Quantum Dissipation}}}\ (\bibinfo  {publisher} {Cambridge University Press},\
  \bibinfo {address} {Cambridge},\ \bibinfo {year} {2014})\BibitemShut
  {NoStop}%
\bibitem [{\citenamefont {Caldeira}\ and\ \citenamefont
  {Leggett}(1981)}]{Caldeira1981}%
  \BibitemOpen
  \bibfield  {author} {\bibinfo {author} {\bibfnamefont {A.~O.}\ \bibnamefont
  {Caldeira}}\ and\ \bibinfo {author} {\bibfnamefont {A.~J.}\ \bibnamefont
  {Leggett}},\ }\href {https://doi.org/10.1103/physrevlett.46.211} {\bibfield
  {journal} {\bibinfo  {journal} {Phys. Rev. Lett.}\ }\textbf {\bibinfo
  {volume} {46}},\ \bibinfo {pages} {211} (\bibinfo {year} {1981})}\BibitemShut
  {NoStop}%
\bibitem [{\citenamefont {Caldeira}\ and\ \citenamefont
  {Leggett}(1983)}]{Caldeira1983}%
  \BibitemOpen
  \bibfield  {author} {\bibinfo {author} {\bibfnamefont {A.}~\bibnamefont
  {Caldeira}}\ and\ \bibinfo {author} {\bibfnamefont {A.}~\bibnamefont
  {Leggett}},\ }\href {https://doi.org/10.1016/0003-4916(83)90202-6} {\bibfield
   {journal} {\bibinfo  {journal} {Ann. Phys.}\ }\textbf {\bibinfo {volume}
  {149}},\ \bibinfo {pages} {374} (\bibinfo {year} {1983})}\BibitemShut
  {NoStop}%
\bibitem [{\citenamefont {Leggett}\ \emph {et~al.}(1987)\citenamefont
  {Leggett}, \citenamefont {Chakravarty}, \citenamefont {Dorsey}, \citenamefont
  {Fisher}, \citenamefont {Garg},\ and\ \citenamefont {Zwerger}}]{Leggett1987}%
  \BibitemOpen
  \bibfield  {author} {\bibinfo {author} {\bibfnamefont {A.~J.}\ \bibnamefont
  {Leggett}}, \bibinfo {author} {\bibfnamefont {S.}~\bibnamefont
  {Chakravarty}}, \bibinfo {author} {\bibfnamefont {A.~T.}\ \bibnamefont
  {Dorsey}}, \bibinfo {author} {\bibfnamefont {M.~P.~A.}\ \bibnamefont
  {Fisher}}, \bibinfo {author} {\bibfnamefont {A.}~\bibnamefont {Garg}},\ and\
  \bibinfo {author} {\bibfnamefont {W.}~\bibnamefont {Zwerger}},\ }\href
  {https://doi.org/10.1103/revmodphys.59.1} {\bibfield  {journal} {\bibinfo
  {journal} {Rev. Mod. Phys.}\ }\textbf {\bibinfo {volume} {59}},\ \bibinfo
  {pages} {1} (\bibinfo {year} {1987})}\BibitemShut {NoStop}%
\bibitem [{\citenamefont {Coleman}(1977)}]{Coleman1977}%
  \BibitemOpen
  \bibfield  {author} {\bibinfo {author} {\bibfnamefont {S.}~\bibnamefont
  {Coleman}},\ }\href {https://doi.org/10.1103/physrevd.15.2929} {\bibfield
  {journal} {\bibinfo  {journal} {Phys. Rev. D}\ }\textbf {\bibinfo {volume}
  {15}},\ \bibinfo {pages} {2929} (\bibinfo {year} {1977})}\BibitemShut
  {NoStop}%
\bibitem [{\citenamefont {Callan}\ and\ \citenamefont
  {Coleman}(1977)}]{Callan1977}%
  \BibitemOpen
  \bibfield  {author} {\bibinfo {author} {\bibfnamefont {C.~G.}\ \bibnamefont
  {Callan}}\ and\ \bibinfo {author} {\bibfnamefont {S.}~\bibnamefont
  {Coleman}},\ }\href {https://doi.org/10.1103/physrevd.16.1762} {\bibfield
  {journal} {\bibinfo  {journal} {Phys. Rev. D}\ }\textbf {\bibinfo {volume}
  {16}},\ \bibinfo {pages} {1762} (\bibinfo {year} {1977})}\BibitemShut
  {NoStop}%
\bibitem [{\citenamefont {Coleman}(1985)}]{Coleman1985}%
  \BibitemOpen
  \bibfield  {author} {\bibinfo {author} {\bibfnamefont {S.}~\bibnamefont
  {Coleman}},\ }\href {https://doi.org/10.1017/cbo9780511565045} {\emph
  {\bibinfo {title} {Aspects of Symmetry}}}\ (\bibinfo  {publisher} {Cambridge
  University Press},\ \bibinfo {address} {Cambridge},\ \bibinfo {year}
  {1985})\BibitemShut {NoStop}%
\bibitem [{\citenamefont {Loss}\ \emph {et~al.}(1992)\citenamefont {Loss},
  \citenamefont {DiVincenzo},\ and\ \citenamefont {Grinstein}}]{Loss1992}%
  \BibitemOpen
  \bibfield  {author} {\bibinfo {author} {\bibfnamefont {D.}~\bibnamefont
  {Loss}}, \bibinfo {author} {\bibfnamefont {D.~P.}\ \bibnamefont
  {DiVincenzo}},\ and\ \bibinfo {author} {\bibfnamefont {G.}~\bibnamefont
  {Grinstein}},\ }\href {https://doi.org/10.1103/physrevlett.69.3232}
  {\bibfield  {journal} {\bibinfo  {journal} {Phys. Rev. Lett.}\ }\textbf
  {\bibinfo {volume} {69}},\ \bibinfo {pages} {3232} (\bibinfo {year}
  {1992})}\BibitemShut {NoStop}%
\bibitem [{\citenamefont {von Delft}\ and\ \citenamefont
  {Henley}(1992)}]{Delft1992}%
  \BibitemOpen
  \bibfield  {author} {\bibinfo {author} {\bibfnamefont {J.}~\bibnamefont {von
  Delft}}\ and\ \bibinfo {author} {\bibfnamefont {C.~L.}\ \bibnamefont
  {Henley}},\ }\href {https://doi.org/10.1103/physrevlett.69.3236} {\bibfield
  {journal} {\bibinfo  {journal} {Phys. Rev. Lett.}\ }\textbf {\bibinfo
  {volume} {69}},\ \bibinfo {pages} {3236} (\bibinfo {year}
  {1992})}\BibitemShut {NoStop}%
\bibitem [{\citenamefont {Braun}\ and\ \citenamefont {Loss}(1996)}]{Braun1996}%
  \BibitemOpen
  \bibfield  {author} {\bibinfo {author} {\bibfnamefont {H.-B.}\ \bibnamefont
  {Braun}}\ and\ \bibinfo {author} {\bibfnamefont {D.}~\bibnamefont {Loss}},\
  }\href {https://doi.org/10.1103/physrevb.53.3237} {\bibfield  {journal}
  {\bibinfo  {journal} {Phys. Rev. B}\ }\textbf {\bibinfo {volume} {53}},\
  \bibinfo {pages} {3237} (\bibinfo {year} {1996})}\BibitemShut {NoStop}%
\bibitem [{\citenamefont {Leuenberger}\ and\ \citenamefont
  {Loss}(2001)}]{Leuenberger2001}%
  \BibitemOpen
  \bibfield  {author} {\bibinfo {author} {\bibfnamefont {M.~N.}\ \bibnamefont
  {Leuenberger}}\ and\ \bibinfo {author} {\bibfnamefont {D.}~\bibnamefont
  {Loss}},\ }\href {https://doi.org/10.1103/physrevb.63.054414} {\bibfield
  {journal} {\bibinfo  {journal} {Phys. Rev. B}\ }\textbf {\bibinfo {volume}
  {63}},\ \bibinfo {pages} {054414} (\bibinfo {year} {2001})}\BibitemShut
  {NoStop}%
\bibitem [{\citenamefont {Affleck}(1986)}]{Affleck1986}%
  \BibitemOpen
  \bibfield  {author} {\bibinfo {author} {\bibfnamefont {I.}~\bibnamefont
  {Affleck}},\ }\href {https://doi.org/10.1103/physrevlett.56.408} {\bibfield
  {journal} {\bibinfo  {journal} {Physical Review Letters}\ }\textbf {\bibinfo
  {volume} {56}},\ \bibinfo {pages} {408} (\bibinfo {year} {1986})}\BibitemShut
  {NoStop}%
\bibitem [{\citenamefont {Read}\ and\ \citenamefont
  {Sachdev}(1989)}]{Read1989}%
  \BibitemOpen
  \bibfield  {author} {\bibinfo {author} {\bibfnamefont {N.}~\bibnamefont
  {Read}}\ and\ \bibinfo {author} {\bibfnamefont {S.}~\bibnamefont {Sachdev}},\
  }\href {https://doi.org/10.1103/physrevlett.62.1694} {\bibfield  {journal}
  {\bibinfo  {journal} {Phys. Rev. Lett.}\ }\textbf {\bibinfo {volume} {62}},\
  \bibinfo {pages} {1694} (\bibinfo {year} {1989})}\BibitemShut {NoStop}%
\bibitem [{\citenamefont {Senthil}\ \emph {et~al.}(2004)\citenamefont
  {Senthil}, \citenamefont {Vishwanath}, \citenamefont {Balents}, \citenamefont
  {Sachdev},\ and\ \citenamefont {Fisher}}]{Senthil2004}%
  \BibitemOpen
  \bibfield  {author} {\bibinfo {author} {\bibfnamefont {T.}~\bibnamefont
  {Senthil}}, \bibinfo {author} {\bibfnamefont {A.}~\bibnamefont {Vishwanath}},
  \bibinfo {author} {\bibfnamefont {L.}~\bibnamefont {Balents}}, \bibinfo
  {author} {\bibfnamefont {S.}~\bibnamefont {Sachdev}},\ and\ \bibinfo {author}
  {\bibfnamefont {M.~P.~A.}\ \bibnamefont {Fisher}},\ }\href
  {https://doi.org/10.1126/science.1091806} {\bibfield  {journal} {\bibinfo
  {journal} {Science}\ }\textbf {\bibinfo {volume} {303}},\ \bibinfo {pages}
  {1490} (\bibinfo {year} {2004})}\BibitemShut {NoStop}%
\bibitem [{\citenamefont {Bray}\ and\ \citenamefont {Moore}(1982)}]{Bray1982}%
  \BibitemOpen
  \bibfield  {author} {\bibinfo {author} {\bibfnamefont {A.~J.}\ \bibnamefont
  {Bray}}\ and\ \bibinfo {author} {\bibfnamefont {M.~A.}\ \bibnamefont
  {Moore}},\ }\href {https://doi.org/10.1103/physrevlett.49.1545} {\bibfield
  {journal} {\bibinfo  {journal} {Phys. Rev. Lett.}\ }\textbf {\bibinfo
  {volume} {49}},\ \bibinfo {pages} {1545} (\bibinfo {year}
  {1982})}\BibitemShut {NoStop}%
\bibitem [{\citenamefont {Schmid}(1983)}]{Schmid1983}%
  \BibitemOpen
  \bibfield  {author} {\bibinfo {author} {\bibfnamefont {A.}~\bibnamefont
  {Schmid}},\ }\href {https://doi.org/10.1103/physrevlett.51.1506} {\bibfield
  {journal} {\bibinfo  {journal} {Phys. Rev. Lett.}\ }\textbf {\bibinfo
  {volume} {51}},\ \bibinfo {pages} {1506} (\bibinfo {year}
  {1983})}\BibitemShut {NoStop}%
\bibitem [{\citenamefont {Callan}\ and\ \citenamefont
  {Freed}(1992)}]{Callan1992}%
  \BibitemOpen
  \bibfield  {author} {\bibinfo {author} {\bibfnamefont {C.~G.}\ \bibnamefont
  {Callan}}\ and\ \bibinfo {author} {\bibfnamefont {D.}~\bibnamefont {Freed}},\
  }\href {https://doi.org/10.1016/0550-3213(92)90400-6} {\bibfield  {journal}
  {\bibinfo  {journal} {Nuclear Physics B}\ }\textbf {\bibinfo {volume}
  {374}},\ \bibinfo {pages} {543} (\bibinfo {year} {1992})}\BibitemShut
  {NoStop}%
\bibitem [{\citenamefont {Garg}\ and\ \citenamefont {Kim}(1989)}]{Garg1989}%
  \BibitemOpen
  \bibfield  {author} {\bibinfo {author} {\bibfnamefont {A.}~\bibnamefont
  {Garg}}\ and\ \bibinfo {author} {\bibfnamefont {G.-H.}\ \bibnamefont {Kim}},\
  }\href {https://doi.org/10.1103/physrevlett.63.2512} {\bibfield  {journal}
  {\bibinfo  {journal} {Phys. Rev. Lett.}\ }\textbf {\bibinfo {volume} {63}},\
  \bibinfo {pages} {2512} (\bibinfo {year} {1989})}\BibitemShut {NoStop}%
\bibitem [{\citenamefont {Garg}(1993{\natexlab{a}})}]{Garg1993a}%
  \BibitemOpen
  \bibfield  {author} {\bibinfo {author} {\bibfnamefont {A.}~\bibnamefont
  {Garg}},\ }\href {https://doi.org/10.1103/physrevlett.70.1541} {\bibfield
  {journal} {\bibinfo  {journal} {Phys. Rev. Lett.}\ }\textbf {\bibinfo
  {volume} {70}},\ \bibinfo {pages} {1541} (\bibinfo {year}
  {1993}{\natexlab{a}})}\BibitemShut {NoStop}%
\bibitem [{\citenamefont {Garg}(1994)}]{Garg1994}%
  \BibitemOpen
  \bibfield  {author} {\bibinfo {author} {\bibfnamefont {A.}~\bibnamefont
  {Garg}},\ }\href {https://doi.org/10.1063/1.358342} {\bibfield  {journal}
  {\bibinfo  {journal} {J. Appl. Phys.}\ }\textbf {\bibinfo {volume} {76}},\
  \bibinfo {pages} {6168} (\bibinfo {year} {1994})}\BibitemShut {NoStop}%
\bibitem [{\citenamefont {Chang}\ and\ \citenamefont
  {Chakravarty}(1984)}]{Chang1984}%
  \BibitemOpen
  \bibfield  {author} {\bibinfo {author} {\bibfnamefont {L.-D.}\ \bibnamefont
  {Chang}}\ and\ \bibinfo {author} {\bibfnamefont {S.}~\bibnamefont
  {Chakravarty}},\ }\href {https://doi.org/10.1103/physrevb.29.130} {\bibfield
  {journal} {\bibinfo  {journal} {Phys. Rev. B}\ }\textbf {\bibinfo {volume}
  {29}},\ \bibinfo {pages} {130} (\bibinfo {year} {1984})}\BibitemShut
  {NoStop}%
\bibitem [{\citenamefont {Grabert}\ \emph {et~al.}(1987)\citenamefont
  {Grabert}, \citenamefont {Olschowski},\ and\ \citenamefont
  {Weiss}}]{Grabert1987}%
  \BibitemOpen
  \bibfield  {author} {\bibinfo {author} {\bibfnamefont {H.}~\bibnamefont
  {Grabert}}, \bibinfo {author} {\bibfnamefont {P.}~\bibnamefont
  {Olschowski}},\ and\ \bibinfo {author} {\bibfnamefont {U.}~\bibnamefont
  {Weiss}},\ }\href {https://doi.org/10.1103/physrevb.36.1931} {\bibfield
  {journal} {\bibinfo  {journal} {Phys. Rev. B}\ }\textbf {\bibinfo {volume}
  {36}},\ \bibinfo {pages} {1931} (\bibinfo {year} {1987})}\BibitemShut
  {NoStop}%
\bibitem [{\citenamefont {Goddard}\ and\ \citenamefont
  {Olive}(1978)}]{Goddard1978}%
  \BibitemOpen
  \bibfield  {author} {\bibinfo {author} {\bibfnamefont {P.}~\bibnamefont
  {Goddard}}\ and\ \bibinfo {author} {\bibfnamefont {D.~I.}\ \bibnamefont
  {Olive}},\ }\href {https://doi.org/10.1088/0034-4885/41/9/001} {\bibfield
  {journal} {\bibinfo  {journal} {Rep. Prog. Phys.}\ }\textbf {\bibinfo
  {volume} {41}},\ \bibinfo {pages} {1357} (\bibinfo {year}
  {1978})}\BibitemShut {NoStop}%
\bibitem [{\citenamefont {Castelnovo}\ \emph {et~al.}(2008)\citenamefont
  {Castelnovo}, \citenamefont {Moessner},\ and\ \citenamefont
  {Sondhi}}]{spinice0}%
  \BibitemOpen
  \bibfield  {author} {\bibinfo {author} {\bibfnamefont {C.}~\bibnamefont
  {Castelnovo}}, \bibinfo {author} {\bibfnamefont {R.}~\bibnamefont
  {Moessner}},\ and\ \bibinfo {author} {\bibfnamefont {S.~L.}\ \bibnamefont
  {Sondhi}},\ }\href {https://doi.org/10.1038/nature06433} {\bibfield
  {journal} {\bibinfo  {journal} {Nature}\ }\textbf {\bibinfo {volume} {451}},\
  \bibinfo {pages} {42} (\bibinfo {year} {2008})}\BibitemShut {NoStop}%
\bibitem [{\citenamefont {Zhang}\ \emph {et~al.}(2016)\citenamefont {Zhang},
  \citenamefont {Mishchenko}, \citenamefont {De~Filippis},\ and\ \citenamefont
  {Nagaosa}}]{XXZ:resistivity}%
  \BibitemOpen
  \bibfield  {author} {\bibinfo {author} {\bibfnamefont {X.-X.}\ \bibnamefont
  {Zhang}}, \bibinfo {author} {\bibfnamefont {A.~S.}\ \bibnamefont
  {Mishchenko}}, \bibinfo {author} {\bibfnamefont {G.}~\bibnamefont
  {De~Filippis}},\ and\ \bibinfo {author} {\bibfnamefont {N.}~\bibnamefont
  {Nagaosa}},\ }\href {https://doi.org/10.1103/PhysRevB.94.174428} {\bibfield
  {journal} {\bibinfo  {journal} {Phys. Rev. B}\ }\textbf {\bibinfo {volume}
  {94}},\ \bibinfo {pages} {174428} (\bibinfo {year} {2016})}\BibitemShut
  {NoStop}%
\bibitem [{\citenamefont {Vafek}\ and\ \citenamefont
  {Vishwanath}(2014)}]{DiracFermion1}%
  \BibitemOpen
  \bibfield  {author} {\bibinfo {author} {\bibfnamefont {O.}~\bibnamefont
  {Vafek}}\ and\ \bibinfo {author} {\bibfnamefont {A.}~\bibnamefont
  {Vishwanath}},\ }\href
  {https://doi.org/10.1146/annurev-conmatphys-031113-133841} {\bibfield
  {journal} {\bibinfo  {journal} {Annual Review of Condensed Matter Physics}\
  }\textbf {\bibinfo {volume} {5}},\ \bibinfo {pages} {83} (\bibinfo {year}
  {2014})}\BibitemShut {NoStop}%
\bibitem [{\citenamefont {Kugel{\textquotesingle}}\ and\ \citenamefont
  {Khomski{\u{\i}}}(1982)}]{Khomskii1982}%
  \BibitemOpen
  \bibfield  {author} {\bibinfo {author} {\bibfnamefont {K.~I.}\ \bibnamefont
  {Kugel{\textquotesingle}}}\ and\ \bibinfo {author} {\bibfnamefont {D.~I.}\
  \bibnamefont {Khomski{\u{\i}}}},\ }\href
  {https://doi.org/10.1070/pu1982v025n04abeh004537} {\bibfield  {journal}
  {\bibinfo  {journal} {Soviet Physics Uspekhi}\ }\textbf {\bibinfo {volume}
  {25}},\ \bibinfo {pages} {231} (\bibinfo {year} {1982})}\BibitemShut
  {NoStop}%
\bibitem [{\citenamefont {Yarkony}(1996)}]{Yarkony1996}%
  \BibitemOpen
  \bibfield  {author} {\bibinfo {author} {\bibfnamefont {D.~R.}\ \bibnamefont
  {Yarkony}},\ }\href {https://doi.org/10.1103/revmodphys.68.985} {\bibfield
  {journal} {\bibinfo  {journal} {Reviews of Modern Physics}\ }\textbf
  {\bibinfo {volume} {68}},\ \bibinfo {pages} {985} (\bibinfo {year}
  {1996})}\BibitemShut {NoStop}%
\bibitem [{\citenamefont {Domcke}\ \emph {et~al.}(2004)\citenamefont {Domcke},
  \citenamefont {Yarkony},\ and\ \citenamefont {Köppel}}]{Domcke2004}%
  \BibitemOpen
  \bibinfo {editor} {\bibfnamefont {W.}~\bibnamefont {Domcke}}, \bibinfo
  {editor} {\bibfnamefont {D.~R.}\ \bibnamefont {Yarkony}},\ and\ \bibinfo
  {editor} {\bibfnamefont {H.}~\bibnamefont {Köppel}},\ eds.,\ \href
  {https://doi.org/10.1142/5406} {\emph {\bibinfo {title} {Conical
  Intersections: Electronic Structure, Dynamics and Spectroscopy}}}\ (\bibinfo
  {publisher} {{World} {Scientific}},\ \bibinfo {address} {Singapore},\
  \bibinfo {year} {2004})\BibitemShut {NoStop}%
\bibitem [{\citenamefont {H\"{a}nggi}\ \emph {et~al.}(1990)\citenamefont
  {H\"{a}nggi}, \citenamefont {Talkner},\ and\ \citenamefont
  {Borkovec}}]{Haenggi1990}%
  \BibitemOpen
  \bibfield  {author} {\bibinfo {author} {\bibfnamefont {P.}~\bibnamefont
  {H\"{a}nggi}}, \bibinfo {author} {\bibfnamefont {P.}~\bibnamefont
  {Talkner}},\ and\ \bibinfo {author} {\bibfnamefont {M.}~\bibnamefont
  {Borkovec}},\ }\href {https://doi.org/10.1103/revmodphys.62.251} {\bibfield
  {journal} {\bibinfo  {journal} {Reviews of Modern Physics}\ }\textbf
  {\bibinfo {volume} {62}},\ \bibinfo {pages} {251} (\bibinfo {year}
  {1990})}\BibitemShut {NoStop}%
\bibitem [{\citenamefont {Kryvohuz}(2011)}]{Kryvohuz2011}%
  \BibitemOpen
  \bibfield  {author} {\bibinfo {author} {\bibfnamefont {M.}~\bibnamefont
  {Kryvohuz}},\ }\href {https://doi.org/10.1063/1.3565425} {\bibfield
  {journal} {\bibinfo  {journal} {The Journal of Chemical Physics}\ }\textbf
  {\bibinfo {volume} {134}},\ \bibinfo {pages} {114103} (\bibinfo {year}
  {2011})}\BibitemShut {NoStop}%
\bibitem [{\citenamefont {Richardson}\ \emph {et~al.}(2011)\citenamefont
  {Richardson}, \citenamefont {Althorpe},\ and\ \citenamefont
  {Wales}}]{Richardson2011}%
  \BibitemOpen
  \bibfield  {author} {\bibinfo {author} {\bibfnamefont {J.~O.}\ \bibnamefont
  {Richardson}}, \bibinfo {author} {\bibfnamefont {S.~C.}\ \bibnamefont
  {Althorpe}},\ and\ \bibinfo {author} {\bibfnamefont {D.~J.}\ \bibnamefont
  {Wales}},\ }\href {https://doi.org/10.1063/1.3640429} {\bibfield  {journal}
  {\bibinfo  {journal} {The Journal of Chemical Physics}\ }\textbf {\bibinfo
  {volume} {135}},\ \bibinfo {pages} {124109} (\bibinfo {year}
  {2011})}\BibitemShut {NoStop}%
\bibitem [{\citenamefont {Child}(2014)}]{Child2014}%
  \BibitemOpen
  \bibfield  {author} {\bibinfo {author} {\bibfnamefont {M.~S.}\ \bibnamefont
  {Child}},\ }\href {https://doi.org/10.1093/acprof:oso/9780199672981.001.0001}
  {\emph {\bibinfo {title} {Semiclassical Mechanics with Molecular
  Applications}}}\ (\bibinfo  {publisher} {Oxford University Press},\ \bibinfo
  {address} {New York},\ \bibinfo {year} {2014})\BibitemShut {NoStop}%
\bibitem [{\citenamefont {Meisner}\ and\ \citenamefont
  {K\"{a}stner}(2016)}]{Meisner2016}%
  \BibitemOpen
  \bibfield  {author} {\bibinfo {author} {\bibfnamefont {J.}~\bibnamefont
  {Meisner}}\ and\ \bibinfo {author} {\bibfnamefont {J.}~\bibnamefont
  {K\"{a}stner}},\ }\href {https://doi.org/10.1002/anie.201511028} {\bibfield
  {journal} {\bibinfo  {journal} {Angewandte Chemie International Edition}\
  }\textbf {\bibinfo {volume} {55}},\ \bibinfo {pages} {5400} (\bibinfo {year}
  {2016})}\BibitemShut {NoStop}%
\bibitem [{\citenamefont {Altland}\ and\ \citenamefont {Simons}(2010)}]{CMFT}%
  \BibitemOpen
  \bibfield  {author} {\bibinfo {author} {\bibfnamefont {A.}~\bibnamefont
  {Altland}}\ and\ \bibinfo {author} {\bibfnamefont {B.~D.}\ \bibnamefont
  {Simons}},\ }\href {http://amazon.com/o/ASIN/0521769752/} {\emph {\bibinfo
  {title} {Condensed Matter Field Theory}}},\ \bibinfo {edition} {2nd}\ ed.\
  (\bibinfo  {publisher} {Cambridge University Press},\ \bibinfo {address}
  {Cambridge},\ \bibinfo {year} {2010})\BibitemShut {NoStop}%
\bibitem [{\citenamefont {Nagaosa}\ \emph {et~al.}(2012)\citenamefont
  {Nagaosa}, \citenamefont {Yu},\ and\ \citenamefont {Tokura}}]{EEMF2}%
  \BibitemOpen
  \bibfield  {author} {\bibinfo {author} {\bibfnamefont {N.}~\bibnamefont
  {Nagaosa}}, \bibinfo {author} {\bibfnamefont {X.~Z.}\ \bibnamefont {Yu}},\
  and\ \bibinfo {author} {\bibfnamefont {Y.}~\bibnamefont {Tokura}},\ }\href
  {https://doi.org/10.1098/rsta.2011.0405} {\bibfield  {journal} {\bibinfo
  {journal} {Philos. Trans. A: Math. Phys. Eng. Sci.}\ }\textbf {\bibinfo
  {volume} {370}},\ \bibinfo {pages} {5806} (\bibinfo {year}
  {2012})}\BibitemShut {NoStop}%
\bibitem [{\citenamefont {Leggett}(1997)}]{Leggett1997}%
  \BibitemOpen
  \bibfield  {author} {\bibinfo {author} {\bibfnamefont {A.~J.}\ \bibnamefont
  {Leggett}},\ }in\ \href {https://doi.org/10.1142/9789812819895_0041} {\emph
  {\bibinfo {booktitle} {Foundations of Quantum Mechanics in the Light of New
  Technology}}}\ (\bibinfo  {publisher} {{World} {Scientific}},\ \bibinfo
  {address} {Singapore},\ \bibinfo {year} {1997})\ pp.\ \bibinfo {pages}
  {406--413}\BibitemShut {NoStop}%
\bibitem [{\citenamefont {Arikoglu}\ and\ \citenamefont
  {Ozkol}(2005)}]{Arikoglu2005}%
  \BibitemOpen
  \bibfield  {author} {\bibinfo {author} {\bibfnamefont {A.}~\bibnamefont
  {Arikoglu}}\ and\ \bibinfo {author} {\bibfnamefont {I.}~\bibnamefont
  {Ozkol}},\ }\href {https://doi.org/10.1016/j.amc.2004.10.009} {\bibfield
  {journal} {\bibinfo  {journal} {Applied Mathematics and Computation}\
  }\textbf {\bibinfo {volume} {168}},\ \bibinfo {pages} {1145} (\bibinfo {year}
  {2005})}\BibitemShut {NoStop}%
\bibitem [{\citenamefont {Trefethen}\ \emph {et~al.}(2017)\citenamefont
  {Trefethen}, \citenamefont {Birkisson},\ and\ \citenamefont
  {Driscoll}}]{Trefethen2017}%
  \BibitemOpen
  \bibfield  {author} {\bibinfo {author} {\bibfnamefont {L.~N.}\ \bibnamefont
  {Trefethen}}, \bibinfo {author} {\bibfnamefont {{\'{A}}.}~\bibnamefont
  {Birkisson}},\ and\ \bibinfo {author} {\bibfnamefont {T.~A.}\ \bibnamefont
  {Driscoll}},\ }\href
  {https://my.siam.org/Store/Product/viewproduct/?ProductId=29219232} {\emph
  {\bibinfo {title} {Exploring ODEs}}}\ (\bibinfo  {publisher} {SIAM-Society
  for Industrial and Applied Mathematics},\ \bibinfo {address} {Philadelphia},\
  \bibinfo {year} {2017})\BibitemShut {NoStop}%
\bibitem [{\citenamefont {He}(2006)}]{He2006}%
  \BibitemOpen
  \bibfield  {author} {\bibinfo {author} {\bibfnamefont {J.-H.}\ \bibnamefont
  {He}},\ }\href {https://doi.org/10.1142/s0217979206033796} {\bibfield
  {journal} {\bibinfo  {journal} {International Journal of Modern Physics B}\
  }\textbf {\bibinfo {volume} {20}},\ \bibinfo {pages} {1141} (\bibinfo {year}
  {2006})}\BibitemShut {NoStop}%
\bibitem [{\citenamefont {LeVeque}(2007)}]{Leveque2007}%
  \BibitemOpen
  \bibfield  {author} {\bibinfo {author} {\bibfnamefont {R.~J.}\ \bibnamefont
  {LeVeque}},\ }\href {https://doi.org/10.1137/1.9780898717839} {\emph
  {\bibinfo {title} {Finite Difference Methods for Ordinary and Partial
  Differential Equations}}}\ (\bibinfo  {publisher} {SIAM-Society for
  Industrial and Applied Mathematics},\ \bibinfo {address} {Philadelphia},\
  \bibinfo {year} {2007})\BibitemShut {NoStop}%
\bibitem [{\citenamefont {Fornberg}(1988)}]{Fornberg1988}%
  \BibitemOpen
  \bibfield  {author} {\bibinfo {author} {\bibfnamefont {B.}~\bibnamefont
  {Fornberg}},\ }\href {https://doi.org/10.1090/s0025-5718-1988-0935077-0}
  {\bibfield  {journal} {\bibinfo  {journal} {Mathematics of Computation}\
  }\textbf {\bibinfo {volume} {51}},\ \bibinfo {pages} {699} (\bibinfo {year}
  {1988})}\BibitemShut {NoStop}%
\bibitem [{\citenamefont {Abramowitz}\ and\ \citenamefont
  {Stegun}(2014)}]{AbramowitzStegun}%
  \BibitemOpen
  \bibfield  {author} {\bibinfo {author} {\bibfnamefont {M.}~\bibnamefont
  {Abramowitz}}\ and\ \bibinfo {author} {\bibfnamefont {I.}~\bibnamefont
  {Stegun}},\ }\href {http://amazon.com/o/ASIN/161427617X/} {\emph {\bibinfo
  {title} {Handbook of Mathematical Functions with Formulas, Graphs, and
  Mathematical Tables}}}\ (\bibinfo  {publisher} {Martino Fine Books},\
  \bibinfo {address} {Eastford, CT},\ \bibinfo {year} {2014})\BibitemShut
  {NoStop}%
\bibitem [{\citenamefont {Press}\ \emph {et~al.}(2007)\citenamefont {Press},
  \citenamefont {Teukolsky}, \citenamefont {Vetterling},\ and\ \citenamefont
  {Flannery}}]{NumericalRecipes}%
  \BibitemOpen
  \bibfield  {author} {\bibinfo {author} {\bibfnamefont {W.~H.}\ \bibnamefont
  {Press}}, \bibinfo {author} {\bibfnamefont {S.~A.}\ \bibnamefont
  {Teukolsky}}, \bibinfo {author} {\bibfnamefont {W.~T.}\ \bibnamefont
  {Vetterling}},\ and\ \bibinfo {author} {\bibfnamefont {B.~P.}\ \bibnamefont
  {Flannery}},\ }\href {http://www.cambridge.org/9780521880688} {\emph
  {\bibinfo {title} {Numerical Recipes: The Art of Scientific Computing}}},\
  \bibinfo {edition} {3rd}\ ed.\ (\bibinfo  {publisher} {Cambridge University
  Press},\ \bibinfo {address} {Cambridge},\ \bibinfo {year} {2007})\BibitemShut
  {NoStop}%
\bibitem [{\citenamefont {Gough}(2009)}]{GSL}%
  \BibitemOpen
  \bibinfo {editor} {\bibfnamefont {B.}~\bibnamefont {Gough}},\ ed.,\ \href
  {http://amazon.com/o/ASIN/0954612078/} {\emph {\bibinfo {title} {GNU
  Scientific Library Reference Manual}}},\ \bibinfo {edition} {3rd}\ ed.\
  (\bibinfo  {publisher} {Network Theory Ltd.},\ \bibinfo {address} {Bristol},\
  \bibinfo {year} {2009})\BibitemShut {NoStop}%
\bibitem [{\citenamefont {Datta}(2005)}]{Datta2005}%
  \BibitemOpen
  \bibfield  {author} {\bibinfo {author} {\bibfnamefont {S.}~\bibnamefont
  {Datta}},\ }\href {https://doi.org/10.1017/cbo9781139164313} {\emph {\bibinfo
  {title} {Quantum Transport: Atom to Transistor}}}\ (\bibinfo  {publisher}
  {Cambridge University Press},\ \bibinfo {address} {Cambridge},\ \bibinfo
  {year} {2005})\BibitemShut {NoStop}%
\bibitem [{\citenamefont {Kamenev}(2009)}]{Kamenev2009}%
  \BibitemOpen
  \bibfield  {author} {\bibinfo {author} {\bibfnamefont {A.}~\bibnamefont
  {Kamenev}},\ }\href {https://doi.org/10.1017/cbo9781139003667} {\emph
  {\bibinfo {title} {Field Theory of Non-Equilibrium Systems}}}\ (\bibinfo
  {publisher} {Cambridge University Press},\ \bibinfo {address} {Cambridge},\
  \bibinfo {year} {2009})\BibitemShut {NoStop}%
\bibitem [{\citenamefont {Schulman}(2005)}]{Schulman2005}%
  \BibitemOpen
  \bibfield  {author} {\bibinfo {author} {\bibfnamefont {L.~S.}\ \bibnamefont
  {Schulman}},\ }\href
  {https://www.ebook.de/de/product/3560298/l_s_schulman_techniques_and_applications_of_path_integration.html}
  {\emph {\bibinfo {title} {Techniques and Applications of Path Integration}}}\
  (\bibinfo  {publisher} {Dover Publications Inc.},\ \bibinfo {address}
  {Mineola, NY},\ \bibinfo {year} {2005})\BibitemShut {NoStop}%
\bibitem [{\citenamefont {Anderson}\ and\ \citenamefont
  {Yuval}(1971)}]{Anderson1971}%
  \BibitemOpen
  \bibfield  {author} {\bibinfo {author} {\bibfnamefont {P.~W.}\ \bibnamefont
  {Anderson}}\ and\ \bibinfo {author} {\bibfnamefont {G.}~\bibnamefont
  {Yuval}},\ }\href {https://doi.org/10.1088/0022-3719/4/5/011} {\bibfield
  {journal} {\bibinfo  {journal} {Journal of Physics C: Solid State Physics}\
  }\textbf {\bibinfo {volume} {4}},\ \bibinfo {pages} {607} (\bibinfo {year}
  {1971})}\BibitemShut {NoStop}%
\bibitem [{\citenamefont {Soluyanov}\ \emph {et~al.}(2015)\citenamefont
  {Soluyanov}, \citenamefont {Gresch}, \citenamefont {Wang}, \citenamefont
  {Wu}, \citenamefont {Troyer}, \citenamefont {Dai},\ and\ \citenamefont
  {Bernevig}}]{Soluyanov2015}%
  \BibitemOpen
  \bibfield  {author} {\bibinfo {author} {\bibfnamefont {A.~A.}\ \bibnamefont
  {Soluyanov}}, \bibinfo {author} {\bibfnamefont {D.}~\bibnamefont {Gresch}},
  \bibinfo {author} {\bibfnamefont {Z.}~\bibnamefont {Wang}}, \bibinfo {author}
  {\bibfnamefont {Q.}~\bibnamefont {Wu}}, \bibinfo {author} {\bibfnamefont
  {M.}~\bibnamefont {Troyer}}, \bibinfo {author} {\bibfnamefont
  {X.}~\bibnamefont {Dai}},\ and\ \bibinfo {author} {\bibfnamefont {B.~A.}\
  \bibnamefont {Bernevig}},\ }\href {https://doi.org/10.1038/nature15768}
  {\bibfield  {journal} {\bibinfo  {journal} {Nature}\ }\textbf {\bibinfo
  {volume} {527}},\ \bibinfo {pages} {495} (\bibinfo {year}
  {2015})}\BibitemShut {NoStop}%
\bibitem [{\citenamefont {Misaki}\ and\ \citenamefont
  {Nagaosa}(2018)}]{Misaki2018}%
  \BibitemOpen
  \bibfield  {author} {\bibinfo {author} {\bibfnamefont {K.}~\bibnamefont
  {Misaki}}\ and\ \bibinfo {author} {\bibfnamefont {N.}~\bibnamefont
  {Nagaosa}},\ }\href {https://doi.org/10.1103/physreve.98.052225} {\bibfield
  {journal} {\bibinfo  {journal} {Physical Review E}\ }\textbf {\bibinfo
  {volume} {98}},\ \bibinfo {pages} {052225} (\bibinfo {year}
  {2018})}\BibitemShut {NoStop}%
\bibitem [{\citenamefont {Golub}\ and\ \citenamefont
  {Welsch}(1969)}]{Golub1969}%
  \BibitemOpen
  \bibfield  {author} {\bibinfo {author} {\bibfnamefont {G.~H.}\ \bibnamefont
  {Golub}}\ and\ \bibinfo {author} {\bibfnamefont {J.~H.}\ \bibnamefont
  {Welsch}},\ }\href {https://doi.org/10.1090/s0025-5718-69-99647-1} {\bibfield
   {journal} {\bibinfo  {journal} {Mathematics of Computation}\ }\textbf
  {\bibinfo {volume} {23}},\ \bibinfo {pages} {221} (\bibinfo {year}
  {1969})}\BibitemShut {NoStop}%
\bibitem [{\citenamefont {Laurie}(1999)}]{Laurie1999}%
  \BibitemOpen
  \bibfield  {author} {\bibinfo {author} {\bibfnamefont {D.~P.}\ \bibnamefont
  {Laurie}},\ }in\ \href {https://doi.org/10.1007/978-3-0348-8685-7_9} {\emph
  {\bibinfo {booktitle} {Applications and Computation of Orthogonal
  Polynomials}}}\ (\bibinfo  {publisher} {Birkhäuser Basel},\ \bibinfo
  {address} {Basel},\ \bibinfo {year} {1999})\ pp.\ \bibinfo {pages}
  {133--144}\BibitemShut {NoStop}%
\bibitem [{\citenamefont {Garg}(1993{\natexlab{b}})}]{Garg1993}%
  \BibitemOpen
  \bibfield  {author} {\bibinfo {author} {\bibfnamefont {A.}~\bibnamefont
  {Garg}},\ }\href {https://doi.org/10.1209/0295-5075/22/3/008} {\bibfield
  {journal} {\bibinfo  {journal} {Europhysics Letters ({EPL})}\ }\textbf
  {\bibinfo {volume} {22}},\ \bibinfo {pages} {205} (\bibinfo {year}
  {1993}{\natexlab{b}})}\BibitemShut {NoStop}%
\end{thebibliography}%
\end{document}